\documentclass[twocolumn,showpacs,amsmath,amssymb,prd]{revtex4}

\usepackage{graphicx}% Include figure files
\usepackage{dcolumn}% Align table columns on decimal point
\usepackage{bm}% bold math
\usepackage{epsf}
\allowdisplaybreaks

\begin{document}

\title{Quasi-equilibrium Binary Black Hole Initial Data for Dynamical Evolutions}

\author{Hwei-Jang Yo$^{1,2}$, James N. Cook$^1$, Stuart L. Shapiro$^{1,3}$,
        and Thomas W. Baumgarte$^{4,1}$}

\affiliation{$^1$Department of Physics,
        University of Illinois at Urbana-Champaign, Urbana, Illinois 61801\\
$^2$Institute of Astronomy and Astrophysics,
        Academia Sinica, Taipei 115, Taiwan, Republic of China\\
$^3$Department of Astronomy \& NCSA,
        University of Illinois at Urbana-Champaign, Urbana, Illinois 61801\\
$^4$Department of Physics and Astronomy, Bowdoin College, Brunswick,
        Maine 04011}

\begin{abstract}
We present a formalism for constructing quasi-equilibrium binary black
hole initial data suitable for numerical evolution.  We construct
quasi-equilibrium models by imposing an approximate helical Killing
symmetry appropriate for quasi-circular orbits.  We use the sum of two
Kerr-Schild metrics as our background metric, thereby improving on
conformally flat backgrounds that do not accommodate rotating black
holes and providing a horizon-penetrating lapse, convenient for
implementing black hole excision.  We set inner boundary conditions at
an excision radius well inside the apparent horizon and construct
these boundary conditions to incorporate the quasi-equilibrium
condition and recover the solution for isolated black holes in the
limit of large separation.  We use our formalism both to generate
initial data for binary black hole evolutions and to construct a crude
quasi-equilibrium, inspiral sequence for binary black holes of fixed
irreducible mass.
\end{abstract}
\maketitle

\section{Introduction}

The coalescence and merger of binary black holes is expected to be one
of the primary sources of gravitational radiation to be detected by
interferometric gravitational wave detectors (including the Laser
Interferometer Gravitational Wave Observatory, LIGO, and the Laser
Interferometer Space Antenna, LISA).  The detection
and interpretation of black hole mergers will be greatly facilitated
by theoretical predictions for the gravitational waveforms produced by
these events.

For large binary separations, post-Newtonian approximations can be used
to model the binary inspiral and gravitational wave emission to
excellent accuracy \cite{PNresult}.  For small binary separations, when
finite size and non-linear effects become more important, it is
expected that numerical relativity simulations will provide the most
accurate models and wave form predictions.  

Constructing numerical models of the binary inspiral typically
proceeds in two steps (see, e.g.~\cite{twBS03} for a recent review).
In the first step, initial data are constructed by solving the
constraint equations of general relativity.  These initial data, which provide
a snapshot of the gravitational fields at a certain instant of time, are
not unique, and certain freely specifiable functions have to be chosen
in accordance with the astrophysical situation at hand (see also the
review \cite{Cook00}).  In the second step, the initial data are evolved
dynamically forward in time, which provides the subsequent binary 
evolution and with it the emitted gravitational wave signal.

To date neither one of these two steps has been solved completely
satisfactorily.  A number of groups have constructed initial data
describing binary black holes in nearly circular orbit
\cite{cook1994,btw00,ggb1,ggb2,pfeifferetal2000,b02,twbb03,abt04}, and there
have been several attempts at dynamical simulations of binary black
holes \cite{b99,BrandCGHLL00,AlcubBBLNST01,BakerBCLT01,btj04}.
While this effort has made significant
progress, the numerical modeling of binary black hole inspiral remains
an unsolved theoretical problem.

Building on the success of the BSSN formulation of the evolution
equations of general relativity \cite{sn95,bs99}, we have recently
developed a code that can stably evolve single, rotating black holes
for arbitrary long times \cite{ybs02}.
This code adopts a simple excision scheme \cite{ab2001} to remove 
the black hole deep interior and its
singularities from the numerical grid.  Implementation of such a scheme
requires a coordinate system that smoothly extends into the black hole
interior (``horizon-penetrating coordinates''), as for example
Kerr-Schild coordinates (see~\cite{mhs99,mm00}).  Our goal is
to use this code for dynamical simulations of binary black hole
systems in corotating coordinates.  This requires initial data that
describe binary black holes in quasi-circular orbit in
horizon-penetrating coordinates.

The first models of binary black holes in quasi-circular orbits
\cite{cook1994,btw00} adopted the Bowen-York decomposition of the
constraint equations \cite{by1980}.  When combined with maximal
slicing and conformal flatness, the momentum constraints can be solved
analytically, and only the Hamiltonian constraint needs to be solved
numerically.  For generalization to spinning black holes (compare
\cite{pfeifferetal2000}) it may be desirable to abandon conformal
flatness.  A recent spectral implementation by \cite{abt04} shows that
in the extreme mass ratio limit this approach does not recover the
Schwarzschild test particle result, which underlines the need for
alternative solutions.  Furthermore, this approach only provides the
initial data for the gravitational fields, and a suitable coordinate
system for a subsequent evolution has yet to be chosen.  Clearly, it
is desirable to choose a rotating coordinate system in which the
binary appears approximately static (i.e.~a coordinate system that is
based on the existence of an approximate helical Killing vector).
Such a coordinate system is constructed in \cite{tbl2003}.  However,
this coordinate system is not horizon-penetrating, since the lapse is
not strictly positive.  This is undesirable for the dynamical
evolution and singularity excision (but see \cite{btj04}; compare
\cite{hecb2003}). 

An alternative approach, \cite{ggb1,ggb2} adopted a conformal
thin-sandwich decomposition of the constraint equations
\cite{Cook00,wilson-mathews-Frontiers,york-1999} instead of the
Bowen-York decomposition.  This approach seems more appealing for the
construction of quasi-equilibrium initial data, since it allows for the
explicit specification of the time derivatives of the conformally
related metric and the trace of the extrinsic curvature (see also
\cite{pct02,dgg02,tbl2003}.)  In addition, this approach automatically
provides a coordinate system that reflects quasi-equilibrium.  In
\cite{ggb1,ggb2}, this decomposition was combined with the
conformal-imaging approach of \cite{cook1994}.  In addition to leading
to some inconsistencies on the black hole throats (compare
\cite{cook02}) this again yields a lapse that is not strictly
positive.  Attempts to combine the thin-sandwich decomposition with a
puncture approach \cite{brandt1997,btw00} fail because of mutually
exclusive requirements between the different methods \cite{hecb2003}.

In this paper we borrow various ideas and approaches from previous
investigators to construct initial data that are better suitable for
evolution with our dynamical evolution evolution code.  In particular,
we adopt the thin-sandwich decomposition of the constraint equations
together with Kerr-Schild background data.  In contrast to
\cite{pct02} we set the time derivative of the trace of the extrinsic
curvature to zero, which we believe will result in data that are
closer to quasi-equilibrium.  On the excision surface we impose a
boundary condition that is derived from requiring that the time
derivative of the conformal factor vanish there.  We impose circular
orbits by setting the ADM mass equal to the Komar mass, which is
equivalent to imposing a relativistic virial theorem \cite{ggb1,ggb2}
(see also \cite{sb02,twbb03}.)

We solve these equations numerically by finite differencing in 
Cartesian coordinates, which leads to
results that are less accurate than those achieved with spectral
methods (compare \cite{pct02}), but better suited for evolutions with
our dynamical code, which also uses finite differencing and 
Cartesian coordinates.  The
accuracy requirements for initial data are much less stringent than
those for constructing accurate quasi-equilibrium sequences, which are
typically determined from small differences between large numbers.  As
a by-product of our calculations, we nevertheless present a crude inspiral,
quasi-equilibrium binary sequence.

The paper is organized as follows.  In Section II we review the basic
equations, boundary conditions and the construction of quasi-circular
orbits.  We present numerical results in Section III, and we discuss
our findings in Section IV.  We also include several Appendices with
specifics of our numerical implementation.

%===================================
\section{Basic Equations}
%===================================
\label{Bf}
%===================================

\subsection{The Thin-Sandwich Equations}
\label{TSE}

We begin by writing the metric in the ADM form
\begin{equation} \label{adm_metric}
ds^2 = - \alpha^2 dt^2 + \gamma_{ij} (dx^i + \beta^i dt)(dx^j + \beta^j dt),
\end{equation}
where $\alpha$ is the lapse function, $\beta^i$ is the shift vector,
and $\gamma_{ij}$ is the spatial metric.  Throughout this paper, Latin
indices are spatial indices and run from 1 to 3, whereas Greek indices
are spacetime indices and run from 0 to 3.

The Einstein equations can then be decomposed into the Hamiltonian
constraint ${\mathcal H}$ and the momentum constraint ${\mathcal
M}_i$
\begin{eqnarray}
   {\mathcal H} &\equiv& R - K_{ij}K^{ij} + K^2 = 0,
   \label{eq:Hamiltonian-const} \\
   {\mathcal M}^i&\equiv& \nabla_j K^{ij} - \nabla^i K = 0, 
   \label{eq:momentum-const}
\end{eqnarray}
and the evolution equations
\begin{eqnarray}
        \partial_t \gamma_{ij} &=& -2\alpha K_{ij} + \nabla_i\beta_j
                + \nabla_j\beta_i.\label{eq:g-evolution}\\
        \partial_t K_{ij} &=& \alpha \left(
                {R}_{ij} - 2K_{i\ell}K^\ell_j + K K_{ij}\right)  
                - \nabla_i\nabla_j\alpha\nonumber\\
                && 
                + \beta^\ell\nabla_\ell K_{ij}
                + K_{i\ell}\nabla_j\beta^\ell
                + K_{j\ell}\nabla_i\beta^\ell ,\label{eq:K-evolution}
\end{eqnarray}
Here we have assumed a vacuum spacetime ($T_{\alpha\beta} = 0$), and
$\nabla_i$, $R_{ij}$ and $R \equiv \gamma^{ij} R_{ij}$ are the
covariant derivative, the Ricci tensor and scalar curvature associated
with the spatial metric $\gamma_{ij}$.  The extrinsic curvature
$K_{ij}$ is defined by equation (\ref{eq:g-evolution}).

Most decompositions of the constraint equations start with a
York--Lichnerowicz conformal decomposition of the metric
\begin{equation}\label{eq:conformal-metric}
        \gamma_{ij} \equiv \psi^4 \tilde\gamma_{ij},
\end{equation}
where $\psi$ is the conformal factor and  $\tilde\gamma_{ij}$ the 
conformally related metric \cite{LichA44,york79}.  It is also useful
to decompose the extrinsic curvature $K_{ij}$ into its trace $K$ and
a tracefree part $A_{ij}$,
\begin{equation}\label{eq:trace-free-K}
        K_{ij} \equiv A_{ij} + {\textstyle\frac13}\gamma_{ij} K,
\end{equation}
and to conformally transform $A_{ij}$ according to
\begin{equation}\label{eq:conf-A}
        A^{ij} \equiv \psi^{-10}\tilde{A}^{ij}
\end{equation}
(so that $A_{ij} = \psi^{-2}\tilde{A}_{ij}$; see
\cite{york-1973,york-1974}.)  With these definitions the Hamiltonian
constraint (\ref{eq:Hamiltonian-const}) becomes
\begin{equation} \label{Ham1}
\tilde\nabla^2\psi - {\textstyle\frac18}\psi\tilde{R}
                - {\textstyle\frac1{12}}\psi^5K^2
                + {\textstyle\frac18}\psi^{-7}\tilde{A}_{ij}\tilde{A}^{ij}
                = 0,
\end{equation}
where $\tilde\nabla_i$ and $\tilde{R}$ are the covariant derivative
and scalar curvature associated with $\tilde\gamma_{ij}$, and
$\tilde\nabla^2 \equiv \tilde\nabla^i\tilde\nabla_i$ is the scalar
Laplace operator.

For a complete derivation of the conformal thin-sandwich decomposition
we refer the reader to references
\cite{wilson-mathews-Frontiers,cook-etal-1996,york-1999}.  Here we
focus on the construction of binary black holes in quasi-equilibrium.
In a corotating coordinate system one expects the gravitational field
to depend on time only very weakly, and it is therefore natural to
construct initial data for which as many functions as possible have
vanishing time derivative.  Within the conformal thin-sandwich
formalism both the time derivative of the conformally related metric
and the extrinsic curvature appear as freely specifiable data, and it
is therefore both possible and natural to set
\begin{equation} \label{Kdot}
\partial_t K = 0
\end{equation}
and
\begin{equation}
\partial_t \tilde \gamma_{ij} = 0.
\end{equation}
Inserting the latter into (\ref{eq:g-evolution}) we obtain
\begin{equation}\label{eq:TS-conf-gdot}
        \tilde{A}^{ij} = \frac{\psi^6}{2\alpha}\left(
                (\tilde{\mathbb L}\beta)^{ij}\right).
\end{equation}
where
\begin{equation}\label{eq:TT-op-def}
 (\tilde{\mathbb L}X)^{ij} \equiv \tilde \nabla^iX^j + \tilde \nabla^jX^i
                - {\textstyle\frac23} \tilde \gamma^{ij} 
		\tilde \nabla_\ell X^\ell.
\end{equation}
Equation (\ref{eq:TS-conf-gdot}) can now be inserted into the Momentum
constraint (\ref{eq:momentum-const}), which yields
\begin{equation}
  \tilde\Delta_{\mathbb L}\beta^i \nonumber
   - (\tilde{\mathbb L}\beta)^{ij}\tilde\nabla_j\ln(\frac{\alpha}{\psi^6})
   = \frac{4}{3}\alpha\tilde\nabla^iK, \label{eq:TS-momentum-const}
\end{equation}
where
\begin{equation}
\tilde\Delta_{\mathbb L}\beta^i=\tilde{\nabla}_j(\tilde{\mathbb L}\beta)^{ij}
      = \tilde{\nabla}^2\beta^i+\frac{1}{3}\tilde{\nabla}^i(\tilde{\nabla}_j
\beta^j)+\tilde{R}^i{}_j\beta^j.
\end{equation}

Finally, condition (\ref{Kdot}) can be inserted into the trace of
the evolution equation (\ref{eq:K-evolution}), which, after combining
with the Hamiltonian constraint (\ref{Ham1}) becomes
\begin{eqnarray}\label{eq:const-trace-K-eqn}
&&\tilde\nabla^2(\alpha\psi) - \alpha\left(
                {\textstyle\frac18}\psi\tilde{R}
                + {\textstyle\frac5{12}}\psi^5K^2\right.\nonumber\\
&&\qquad\qquad\qquad\quad +\left.{\textstyle\frac78}\psi^{-7}\!\tilde{A}_{ij}
                \tilde{A}^{ij}\right) = \psi^5\beta^i\tilde\nabla_{\!i}K,
\end{eqnarray}

To summarize, the thin-sandwich formalism then provides three equations
\begin{eqnarray}\label{esstart}
&&\tilde\Delta_{\mathbb L}\beta^i
                - (\tilde{\mathbb L}\beta)^{ij}\tilde\nabla_j\ln
                  \left(\frac{\alpha}{\psi^6}\right)
                - {\textstyle\frac43}\alpha\tilde\nabla^iK = 0 \\
&&\tilde\nabla^2\psi - {\textstyle\frac18}\psi\tilde{R}
                - {\textstyle\frac1{12}}\psi^5K^2
                + {\textstyle\frac18}\psi^{-7}\tilde{A}_{ij}\tilde{A}^{ij} = 0
\label{esmiddle}\\
&&\tilde\nabla^2(\alpha\psi) - (\alpha\psi)\left[
                {\textstyle\frac18}\tilde{R}
                + {\textstyle\frac5{12}}\psi^4K^2\right.\nonumber\\
&&\qquad\qquad\qquad\quad +\left.{\textstyle\frac78}\psi^{-8}\!\tilde{A}_{ij}
                \tilde{A}^{ij}\right] = \psi^5\beta^i\tilde\nabla_{\!i}K,
\label{esend}
\end{eqnarray}
for the three unknowns $\alpha$, $\beta^i$ and $\psi$.  The tracefree
part of the extrinsic curvature is related to these unknowns through 
equation (\ref{eq:TS-conf-gdot}).  Before the equations can be solved,
a background geometry $\tilde \gamma_{ij}$ and a background trace
of the extrinsic curvature $K$ has to be chosen.

\subsection{Kerr-Schild Background Data}
\label{KSbackground}

We base our choice for the freely specifiable data on a superposition of two
Kerr black holes in Kerr-Schild coordinates \cite{mhs99,mm00,pct02}.
A Kerr-Schild metric is given by
\begin{equation}\label{eq:KerrSchild}
  g_{\mu\nu}=\eta_{\mu\nu}+2Hl_\mu l_\nu,
\end{equation}
where $\eta_{\mu\nu}$ is the Minkowski metric, and $l_\mu$ is a
null-vector with respect to both the full metric and the Minkowski
metric, $g^{\mu\nu}l_\mu l_\nu=\eta^{\mu\nu}l_\mu l_\nu=0$.  From the
spacetime metric (\ref{eq:KerrSchild}) the spatial metric, the lapse
and the shift can be identified as
\begin{align}
  \label{eq:KerrSchild-gamma}
  \gamma_{ij}&=\delta_{ij}+2Hl_il_j,\\
  \alpha&=(1+2Hl^tl^t)^{-1/2},\\
\label{eq:KerrSchild-beta}
  \beta^i&=-\frac{2Hl^tl^i}{1+2Hl^tl^t}.
\end{align}
For a black hole of mass $M$ and angular momentum $M\vec a$ at rest at
the origin, $H$ and $l_{\mu}$ are given by
\begin{align}
  H&=\frac{Mr^3}{r^4+(\vec a\cdot\vec x)^2},\\
  l_\mu&=(1, \vec l),\\
  \vec l\;\,&=\frac{r\vec x-\vec a\times\vec x+(\vec
 a\cdot\vec x)\vec a/r}
              {r^2+a^2},
\end{align}
with
\begin{equation}
  r^2=\frac{\vec x^2-\vec a^2}{2}
       +\left(\frac{(\vec x^2- \vec a^2)^2}{4}
       +(\vec a\cdot\vec x)^2\right)^{1/2}.
\end{equation}
For a non-rotating black hole with $\vec a=0$, $H$ has a pole at the
origin, whereas for rotating black holes $H$ has a ring singularity.
We therefore have to excise from the computational domain a region
enclosing the singularity.  In this paper we adopt a non-spinning
Kerr-Schild background to describe co-rotating black hole binaries in
a co-rotating frame.

We want to generate initial data for a spacetime containing two black
holes with background masses $M_A$ and $M_B$, velocities $\vec v_A$
and $\vec v_B$, and we will assume that the background metric has zero
spin $M\vec a$.  Such initial data can be constructed by adopting for
the freely specifiable background data a superposition of two
Kerr-Schild coordinate systems describing two individual black holes
\cite{mhs99,mm00}.  The first black hole with label A has has a
spatial metric
\begin{equation}
  \label{eq:KerrSchild-HoleA-gamma}
\gamma_{A\,ij}=\delta_{ij}+2H_{\!A}\,l_{A\,i}\,l_{A\,j},
\end{equation}
an extrinsic curvature $K_{\!A\,ij}$, a lapse $\alpha_{\!A}$
and a shift $\beta^i_A$. The trace of the extrinsic curvature is
\begin{equation}
K_A=\frac{2M_A}{r^2_A(1+2M_A/r_A)^{3/2}}(1+3M_A/r_A).
\end{equation}
The second black hole has a similar set of associated
quantities which are labeled with the letter B.

In Section \ref{TSE} we have already adopted $\partial_t \tilde
\gamma_{ij} = 0$ and $\partial_t K = 0$, which leaves as the 
freely specifiable background quantities the
background metric $\tilde \gamma_{ij}$ and the trace of the extrinsic
curvature $K$, for which we choose the ``superpositions''
\begin{equation}
  \label{eq:BinaryKerrSchild-gamma}
  \tilde\gamma_{ij}=\delta_{ij}+2H_{\!A}\,l_{A\,i}\,l_{A\,j}
  +2H_{\!B}\,l_{B\,i}\,l_{B\,j}
\end{equation}
and
\begin{equation}
K=K_A + K_B\label{eq:BinaryKerrSchild-K}.
\end{equation}

\subsection{Outer Boundary Conditions}
\label{OBC}

The requirement of corotation and the conditions of asymptotic flatness
yield boundary conditions at spatial infinity
\begin{eqnarray}\label{bcstart}
\psi|_{r\rightarrow\infty}&=&1,\\
\beta^i|_{r\rightarrow\infty}&=&\Omega(\partial_\phi)^i,\\
\alpha|_{r\rightarrow\infty}&=&1,\label{bcend}
\end{eqnarray}
where $\Omega$ is the angular velocity of the corotating frame.  Since
our computational domain does not extend to spatial infinity, we have
to impose approximate boundary conditions at a finite separation.  The
asymptotic behavior of the metric in a binary black hole system is
similar to that of any rotating system in an asymptotically flat
spacetime, including a Kerr black hole.  The asymptotic form of
a Kerr black hole tells us the form of the leading-order, radial
fall-off term of the shift that is important for determining the
system's angular momentum (via a quadrature over the extrinsic
curvature in eqn.~\ref{Ang_i} below).  To incorporate the angular
momentum of the binary in the outer boundary condition of the shift
vector, we consider the asymptotic shift of a single, {\it rotating}
Kerr-Schild black hole and focus on the leading terms proportional to
the spin. In our asymptotic binary shift we ``correct'' the shift
associated with the non-spinning background metric with terms that
have the same asymptotic fall-off as these spin-dependent terms.
A similar argument for choosing the form of the asymptotic shift in our binary
has been put forward in Sec.~III E of \cite{bmnm2003}. 

For a single Kerr-Schild black hole we have
\begin{eqnarray}
%\alpha & = & 1-\frac{M}{r}+\frac{3M^2 }{2r^2}+O(r^{-3}) ,\nonumber \\
\beta^x & = & \frac{2Mx}{r^2}
   +2M \frac{-2Mx+ay}{r^3}+O(r^{-3}), \nonumber \\
\beta^y & = & \frac{2My}{r^2}
   +2M \frac{-2My-ax}{r^3}+O(r^{-3}), \label{expand} \\
\beta^z & = & \frac{2Mz}{r^2}
   -4M^2\frac{z}{r^3} + 2M\frac{(4M^2+a^2)z}{r^4}+O(r^{-4}),
	\nonumber
\end{eqnarray}
where we have assumed rotation about the $z$-axis, $\vec a = (0,0,a)$.
Equation (\ref{expand}) suggests the following fall-off conditions at
the edge of our computational grid for the binary:
\begin{eqnarray}\label{bc2start}
\psi - 1&\sim&\frac{1}{r},\\
\beta^x - \bar{\beta}^x &\sim& \frac{y}{r^3},\label{bbc2start}\\
\beta^y - \bar{\beta}^y &\sim& \frac{x}{r^3},\\
\beta^z - \bar{\beta}^z &\sim& \frac{z}{r^4},\label{bbc2end}\\
\alpha - 1 &\sim& \frac{1}{r}.\label{bc2end}
\end{eqnarray}
Here the boundary condition of the shift vector consists of an
analytic part, $\bar{\beta}^i$, and a higher-order
part. $\bar{\beta}^i$ is the sum of the analytic shifts from each
nonspinning black hole ($a=0$) plus $(\vec \Omega \times \vec r)^i$,
which accounts for all shift terms except for terms due
to the orbital rotation (frame-dragging) as identified above.
The {\it form} of the
higher-order terms on the right-hand side comes from consideration
of the way in which the
system's total angular momentum is embedded in the asymptotic shift,
as we argued above. The {\it coefficients} for the higher-order terms are
determined by numerically fitting to the data immediately interior to
the boundary; they are not given a priori.  Note that the shift in Sec
IV A of \cite{ggb1} exhibits the same asymptotic behavior as in
eqn.~(\ref{bbc2start})-(\ref{bbc2end}), including the higher-order,
fall-off terms containing the angular momentum data.  The boundary
conditions of \cite{ggb1} and ours differ only in the analytical part:
the background shift in \cite{ggb1} is based on isotropic coordinates,
while in our approach it is based on Kerr-Schild coordinates.  Apart
from the background, the form for the leading-order terms, including
the rotational terms, is the same in both calculations (compare eqns
\ref{bbc2start}-\ref{bbc2end} in our paper with eqns 161-163
in \cite{ggb1}).

\subsection{Inner Boundary Conditions}
\label{InnerBC}

Since the metric is singular at the center of each hole, some part of
the black hole interior has to be excised from the computational
domain, which introduces the need for inner boundary conditions.
In \cite{pct02} the following set of inner boundary conditions 
was adopted
\begin{subequations}\label{eq:BC-sandwich}
\begin{align}
\psi&=1                            &&\mbox{all boundaries,}\\
\label{eq:Sandwich-BC2}
\beta^i&=\beta_A^i                 &&\mbox{sphere inside hole A,}\\
\beta^i&=\beta_B^i                 &&\mbox{sphere inside hole B.}
\end{align} 
\end{subequations}
Since \cite{pct02} specified the lapse as either $\alpha =
\psi^6\alpha_{\rm A}\alpha_{\rm B}$ or $\alpha = \psi^6(\alpha_{\rm
A}+\alpha_{\rm B}-1)$, no boundary condition for the lapse was
required.  We solve equation (\ref{eq:const-trace-K-eqn}) for the
lapse, and therefore need an addition boundary condition.

The set of conditions (\ref{eq:BC-sandwich}) is very simple to
implement, but does not necessarily lead to quasi-equilibrium
solutions.  Assuming that the black holes are equilibrium (or
``isolated'' in the language of \cite{abdfklw00}) Cook \cite{cook02}
derived an alternative set of boundary conditions (see also
\cite{Pfeiffer03,cook03}).  Unfortunately, the resulting equations are
quite complicated and difficult to implement numerically.  We have
therefore chosen to adopt an alternative set of boundary conditions,
which is motivated by the observation that for corotating,
quasi-equilibrium black holes in a binary black hole system the time
derivative
\begin{equation}\label{detg_t}
 \partial_t\ln\sqrt{\gamma} = \nabla_i\beta^i - \alpha K
\end{equation}
should be small \cite{JGM04}.

For the lapse and shift we set
\begin{eqnarray}
  \alpha&=&\alpha_{\rm A}\alpha_{\rm B}\\
  \beta^i&=&\alpha_{\rm A}\beta^i_{\rm B}+\alpha_{\rm B}\beta^i_{\rm A}
\end{eqnarray}
on the inner boundaries and note that these choices reduce to the
correct values in the limit of infinite binary separation.  Imposing
inner boundary conditions somewhere inside the black hole horizon
implicitly assumes that the solution should not depend on where
exactly this condition is imposed.  This suggest that the above
condition should hold not only on the boundary itself, but, to a
certain approximation, also in the neighborhood of the boundary.  With
our choices of the conformal metric
\begin{equation}
\gamma_{ij} = \psi^4\tilde{\gamma}_{ij}
= \psi^4 (\gamma_{ij}^{\rm A} + \gamma_{ij}^{\rm B} -\delta_{ij})
\end{equation}
and the trace of the extrinsic curvature
\begin{equation}
K = K_{\rm A} + K_{\rm B}.
\end{equation}
we can then compute
\begin{eqnarray}\label{bbhg_t}
&&\nabla_i\beta^i - \alpha K\nonumber\\
&&=\nabla_i(\alpha_{\rm A}\beta^i_{\rm B}+\alpha_{\rm B}\beta^i_{\rm A})
- \alpha_{\rm A}\alpha_{\rm B}(K_{\rm A} + K_{\rm B})\nonumber\\
&&=\partial_i(\alpha_{\rm A}\beta^i_{\rm B}+\alpha_{\rm B}\beta^i_{\rm A})
  +\Gamma^j{}_{ji}(\alpha_{\rm A}\beta^i_{\rm B}+\alpha_{\rm B}\beta^i_{\rm A})
  \nonumber\\
 &&\quad - \alpha_{\rm A}\alpha_{\rm B}(K_{\rm A} + K_{\rm B})\nonumber\\
&&=\alpha_{\rm A}(\nabla_i^{\rm B}\beta^i_{\rm B} - \alpha_{\rm B}K_{\rm B})
  +\alpha_{\rm B}(\nabla_i^{\rm A}\beta^i_{\rm A} - \alpha_{\rm A}K_{\rm A})
  \nonumber\\
&&\quad+\beta^i_{\rm B}[\partial_i\alpha_{\rm A}+\alpha_{\rm A}(\Gamma^j{}_{ji}
  -{}_{\rm B}\Gamma^j{}_{ji})]\nonumber\\
&&\quad+\beta^i_{\rm A}[\partial_i\alpha_{\rm B}+\alpha_{\rm B}(\Gamma^j{}_{ji}
  -{}_{\rm A}\Gamma^j{}_{ji})]\nonumber\\
&&=\beta^i_{\rm B}[\partial_i\alpha_{\rm A}+\alpha_{\rm A}\partial_i
\ln\sqrt{\gamma/\gamma_{\rm B}}]\nonumber\\
&&\quad+\beta^i_{\rm A}[\partial_i\alpha_{\rm B}+\alpha_{\rm B}\partial_i
\ln\sqrt{\gamma/\gamma_{\rm A}}]\nonumber\\
&&=\beta^i_{\rm B}\alpha_{\rm A}\partial_i\ln\sqrt{\gamma/\gamma_{\rm A}
  \gamma_{\rm B}}
 +\beta^i_{\rm A}\alpha_{\rm B}\partial_i\ln\sqrt{\gamma/\gamma_{\rm A}
  \gamma_{\rm B}}\nonumber\\
&&=\beta^i\partial_i\ln(\sqrt{\gamma}\alpha).
\end{eqnarray}
Here we have used $\nabla_i \beta^i - \alpha K = 0$ as well as
$\Gamma^j{}_{ji}=\partial_i\ln\sqrt{\gamma}$, 
$\tilde{\Gamma}^j{}_{ji}=\partial_i\ln\sqrt{\tilde{\gamma}}$, 
and $\alpha =\gamma^{-1/2}$ 
for both background black holes $A$ and $B$.  With
$\gamma = \psi^{12}\tilde{\gamma}$, equation (\ref{bbhg_t}) can be
rewritten
\begin{equation}
\nabla_i\beta^i - \alpha K=\beta^i\partial_i\ln(\sqrt{\gamma}\alpha)
=\beta^i\partial_i\ln(\alpha\psi^6\sqrt{\tilde{\gamma}}).
\end{equation}
According to (\ref{detg_t}) we expect $\nabla_i\beta^i - \alpha K$ to be
small, so that (\ref{bbhg_t}) becomes
\begin{equation}
6\beta^i\partial_i\psi+\psi\beta^i(\tilde{\Gamma}^j{}_{ji}+\partial_i\ln
\alpha) = 0,
\end{equation}
which provides a Neumann condition for the conformal factor on the inner
boundary.  Collecting the inner boundary conditions, we then have
\begin{eqnarray}
\alpha & = & \alpha_{\rm A}\alpha_{\rm B},\nonumber\\
\beta^i & = & \alpha_{\rm A}\beta^i_{\rm B}+\alpha_{\rm B}\beta^i_{\rm A},
\label{iibc}\\
6\beta^i\partial_i\psi & = &  - 
\psi\beta^i(\tilde{\Gamma}^j{}_{ji}+\partial_i\ln \alpha) \nonumber
\end{eqnarray}
In the limit of infinite separation each black hole reduces to an
isolated Kerr-Schild black hole, which satisfies the above conditions.

\unitlength 1mm
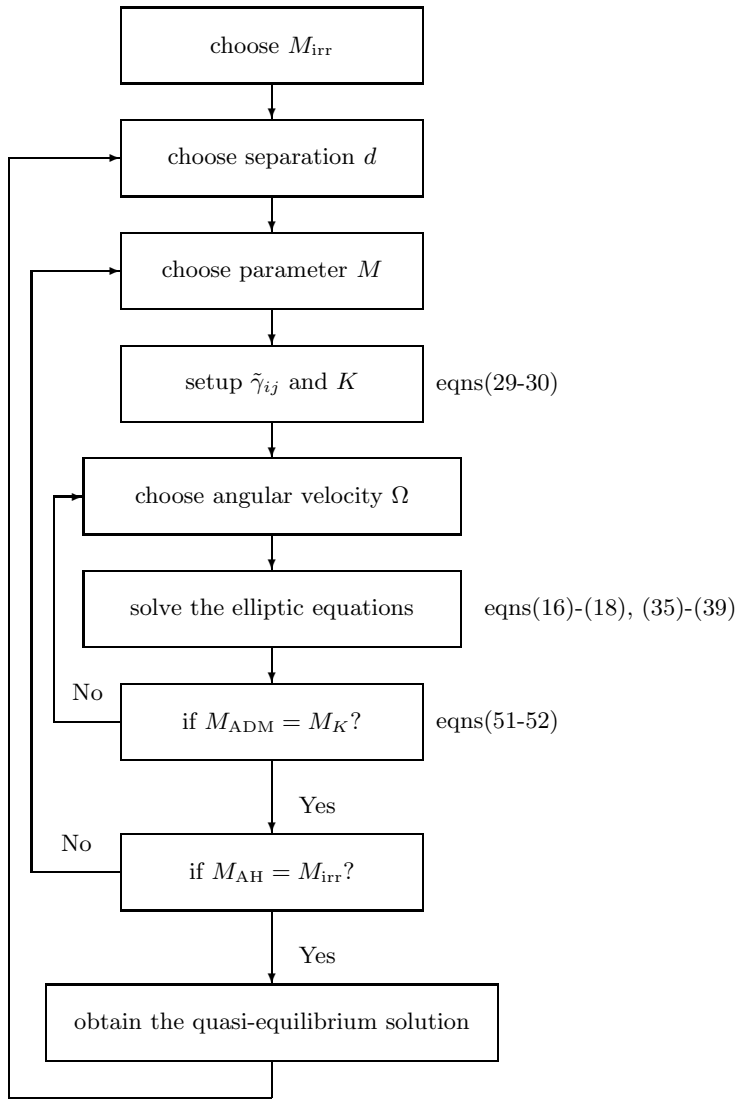
\begin{figure*}
\begin{picture}(70,150)
\put(20,140){\framebox(40,10){choose $M_{\rm irr}$}}
\put(20,125){\framebox(40,10){choose separation $d$}}
\put(20,110){\framebox(40,10){choose parameter $M$}}
\put(20,95){\framebox(40,10){setup $\tilde{\gamma}_{ij}$ and $K$}}
\put(15,80){\framebox(50,10){choose angular velocity $\Omega$}}
\put(15,65){\framebox(50,10){solve the elliptic equations}}
\put(20,50){\framebox(40,10){if $M_{\rm ADM}=M_K$?}}
\put(20,30){\framebox(40,10){if $M_{\rm AH}=M_{\rm irr}$?}}
\put(10,10){\framebox(60,10){obtain the quasi-equilibrium solution}}
\put(40,140){\vector(0,-1){4.8}}
\put(40,125){\vector(0,-1){4.8}}
\put(40,110){\vector(0,-1){4.8}}
\put(40,95){\vector(0,-1){4.8}}
\put(40,80){\vector(0,-1){4.8}}
\put(40,65){\vector(0,-1){4.8}}
\put(40,50){\vector(0,-1){9.8}}
\put(40,30){\vector(0,-1){9.8}}
%%%
\put(40,10){\line(0,-1){5.0}}
\put(40,5){\line(-1,0){35.0}}
\put(5,5){\line(0,1){125.0}}
\put(5,130){\vector(1,0){14.8}}
%%%
\put(20,35){\line(-1,0){12.0}}
\put(8,35){\line(0,1){80.0}}
\put(8,115){\vector(1,0){11.8}}
%%%
\put(20,55){\line(-1,0){9.0}}
\put(11,55){\line(0,1){30.0}}
\put(11,85){\vector(1,0){3.8}}
%%%
\put(11,55){\makebox(9,8){No}}
\put(8,35){\makebox(12,8){No}}
\put(40,20){\makebox(12,8){Yes}}
\put(40,40){\makebox(12,8){Yes}}
%%%
\put(60,50){\makebox(20,10)
{eqns(\ref{eq:ADM-mass-gen-def}-\ref{eq:komar-mass-gen-def})}}
\put(75,65){\makebox(20,10){eqns(\ref{esstart})-(\ref{esend}),
(\ref{bc2start})-(\ref{bc2end})}}
\put(60,95){\makebox(20,10)
{eqns(\ref{eq:BinaryKerrSchild-gamma}-\ref{eq:BinaryKerrSchild-K})}}
\end{picture}
\caption{Flow chart for the construction of sequences of
quasi-equilibrium, circular-orbit binary black holes}
\label{chart}
\end{figure*}

\subsection{Constructing quasi-equilibrium circular orbits and sequences}
\label{sec:obta-circ-orbits}

Solving equations (\ref{esstart}) -- (\ref{esend}) subject to the
boundary conditions (\ref{bc2start}) -- (\ref{bc2end}) and
(\ref{iibc}) yields a solution describing two black holes at a particular
separation $d$, mass $M$ and angular momentum $J$.
Sequences of constant irreducible mass binaries in circular orbit can
be constructed as follows (see also the flow chart in
Fig.~\ref{chart}).

Focusing on equal-mass binaries, we first choose a value of the
irreducible mass $M_{\rm irr}$ \cite{ChristD70,BekenJ73,MTW}, which
remains constant during the slow, adiabatic binary inspiral (see also
\cite{alvi01,ggb2}).  The irreducible mass is determined from the area of the
black hole event horizon, but in practice we approximate this value by
computing the area of the apparent horizon
\begin{equation}
M_{\rm irr} \approx \left( \frac{A}{16 \pi} \right)^{1/2}.
\end{equation}
We next choose a separation $d$, and begin the
iteration with a trial value of the background masses $M_A = M_B = M$,
which enters the background geometry $\tilde \gamma_{ij}$ and $K$.  We
also choose a trial value of the orbital angular velocity $\Omega$,
which enters the orbital shift in $\bar \beta^i$ in the outer boundary
conditions (\ref{bc2start}) -- (\ref{bc2end}).  Solving equations
(\ref{esstart}) -- (\ref{esend}) for these values will provide a
binary that is not necessarily in circular orbit and does not
necessarily have the required irreducible mass.

To impose circular orbits, we require that the system's ADM mass
(e.g.~\cite{murchadha74})
\begin{equation}\label{eq:ADM-mass-gen-def}
        M_{\rm ADM} = \frac1{16\pi}\oint_\infty
           \gamma^{im}\gamma^{jn}(\gamma_{mn,j}-\gamma_{jn,m})
                {\rm d}^2S_i,
\end{equation}
be equal to its Komar mass \cite{komar59}
\begin{equation}\label{eq:komar-mass-gen-def}
        M_{\rm K} = \frac1{4\pi}\oint_\infty{\gamma^{ij}
                (\nabla_i\alpha - \beta^kK_{ik}){\rm d}^2S_j}.
\end{equation}
In the above expressions ${\rm d}^2S_i = (1/2) \gamma^{1/2}
\epsilon_{ijk}dx^jdx^k$ is the two-dimensional surface area element.
In many cases, $\beta^kK_{ik}$ falls off faster than $O(r^{-2})$ in
(\ref{eq:komar-mass-gen-def}) so that the second term vanishes; in our case, 
however, this term cannot be neglected.  We
evaluate these integrals as described in Appendices \ref{ADMmass} and
\ref{KomarMass}.

The equality of the ADM and Komar masses is closely related to a
relativistic virial theorem \cite{gebs94} and indicates that the
spacetime is stationary \cite{beig78,ashtekar79} in the rotating
frame.  In \cite{ggb2} this criterion was adopted to impose circular
orbits in binary black hole spacetimes (see also \cite{sb02} for a
pedagogical illustration).  In our code we iterate over $\Omega$ until
$M_{\rm ADM} = M_{\rm K}$ has been achieved to within an accuracy of
$1$ part in $10^6$.

For a given circular orbit we then iterate over the background mass
$M$ until the irreducible mass $M_{\rm irr}$ has taken the desired
value to within an accuracy of $1$ part in $10^6$ \cite{NoScale}.
Finally we vary the binary separation $d$ to construct an approximate
inspiral sequence.  For each model the ADM mass $M_{\rm ADM}$ is found
from (\ref{eq:ADM-mass-gen-def}) and the angular momentum from
\begin{equation}\label{Ang_i}
%J_i=\frac{\epsilon_{ijk}}{8\pi}\oint_\infty x^j\tilde{A}^{k\ell}d^2S_\ell
J_i=\frac{1}{8\pi}\epsilon_{ij}{}^k\oint_\infty x^jK^\ell{}_kd^2S_\ell
\end{equation}
(see Appendix \ref{angmom}).  The innermost stable circular orbit
(ISCO) can be found by locating minima in the ADM mass and the angular
momentum along a constant-mass sequence.

\section{Numerical Results}

\subsection{Tests}

\begin{figure*}
\begin{tabular}{rl}
\includegraphics[height=7cm,width=8cm]{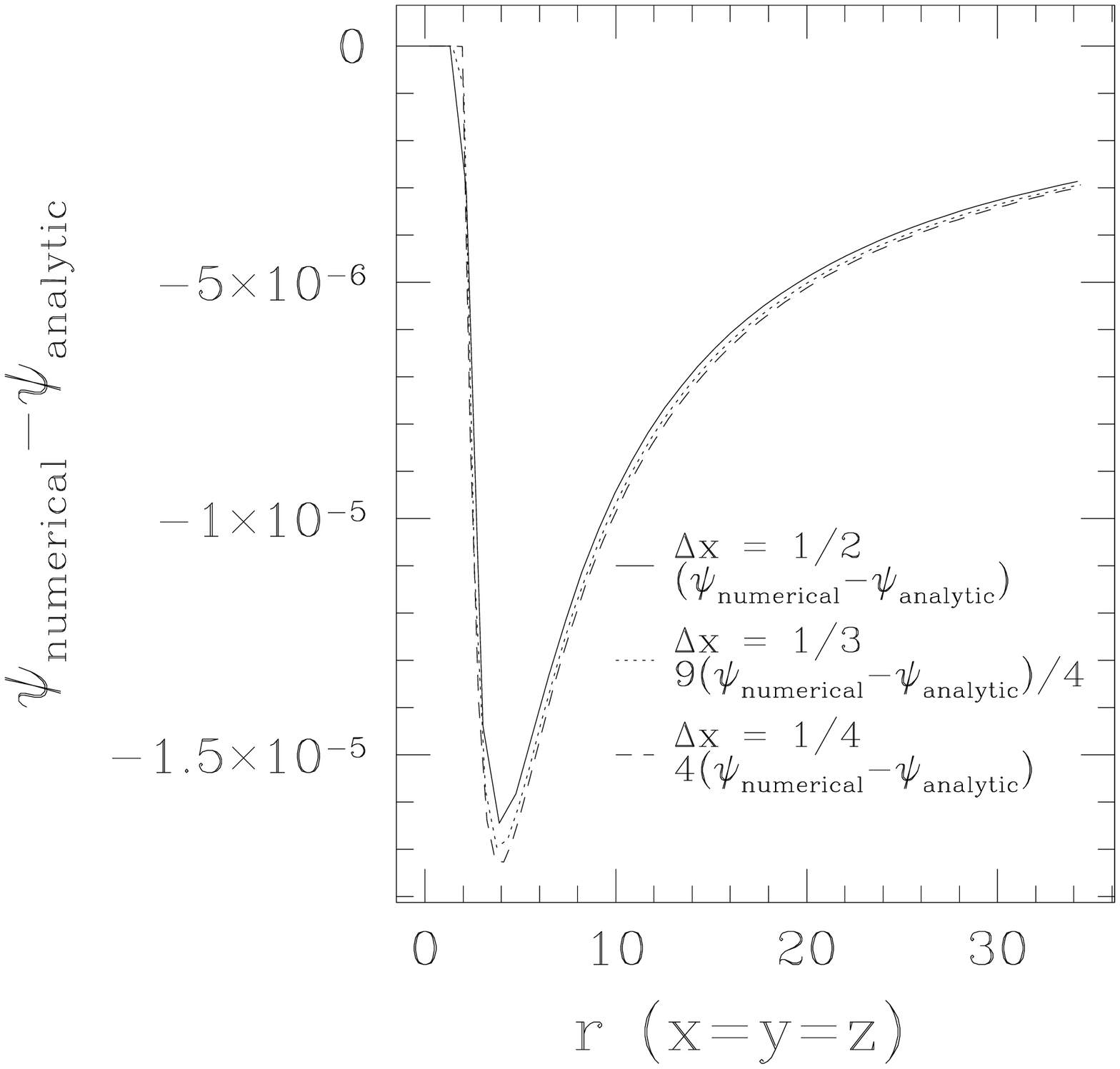}&
\includegraphics[height=7cm,width=8cm]{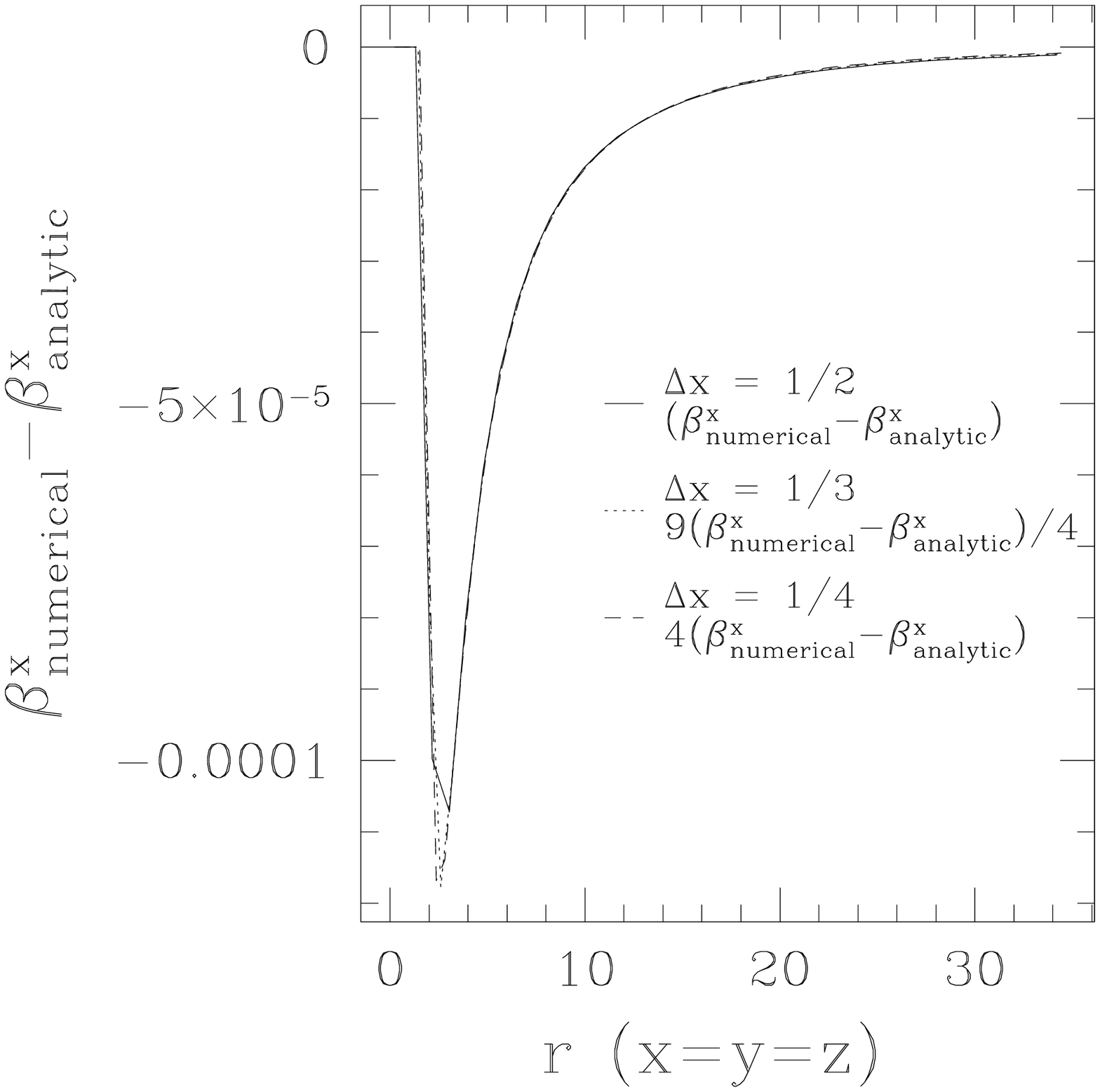} \\
\includegraphics[height=7cm,width=8cm]{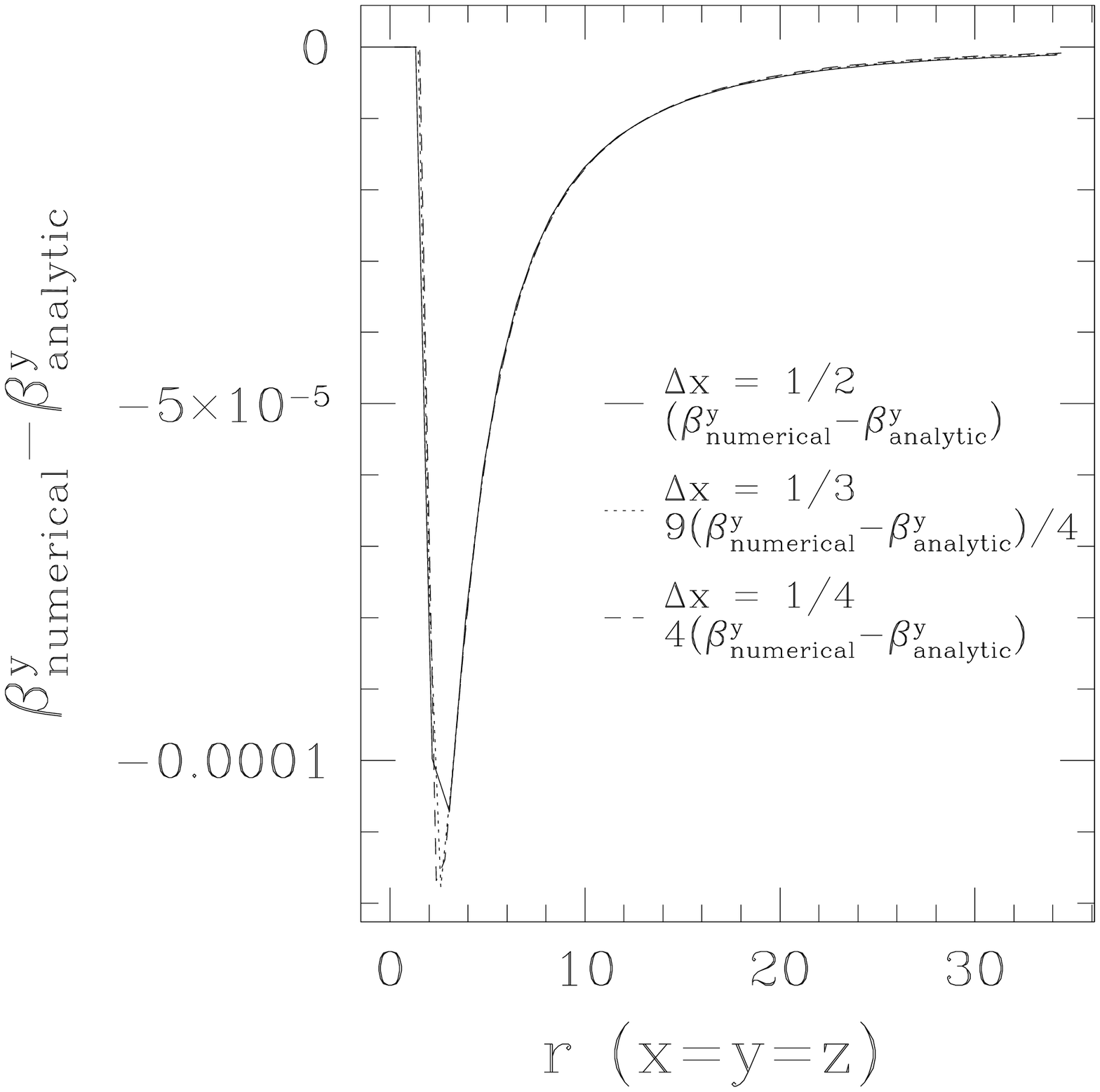}&
\includegraphics[height=7cm,width=8cm]{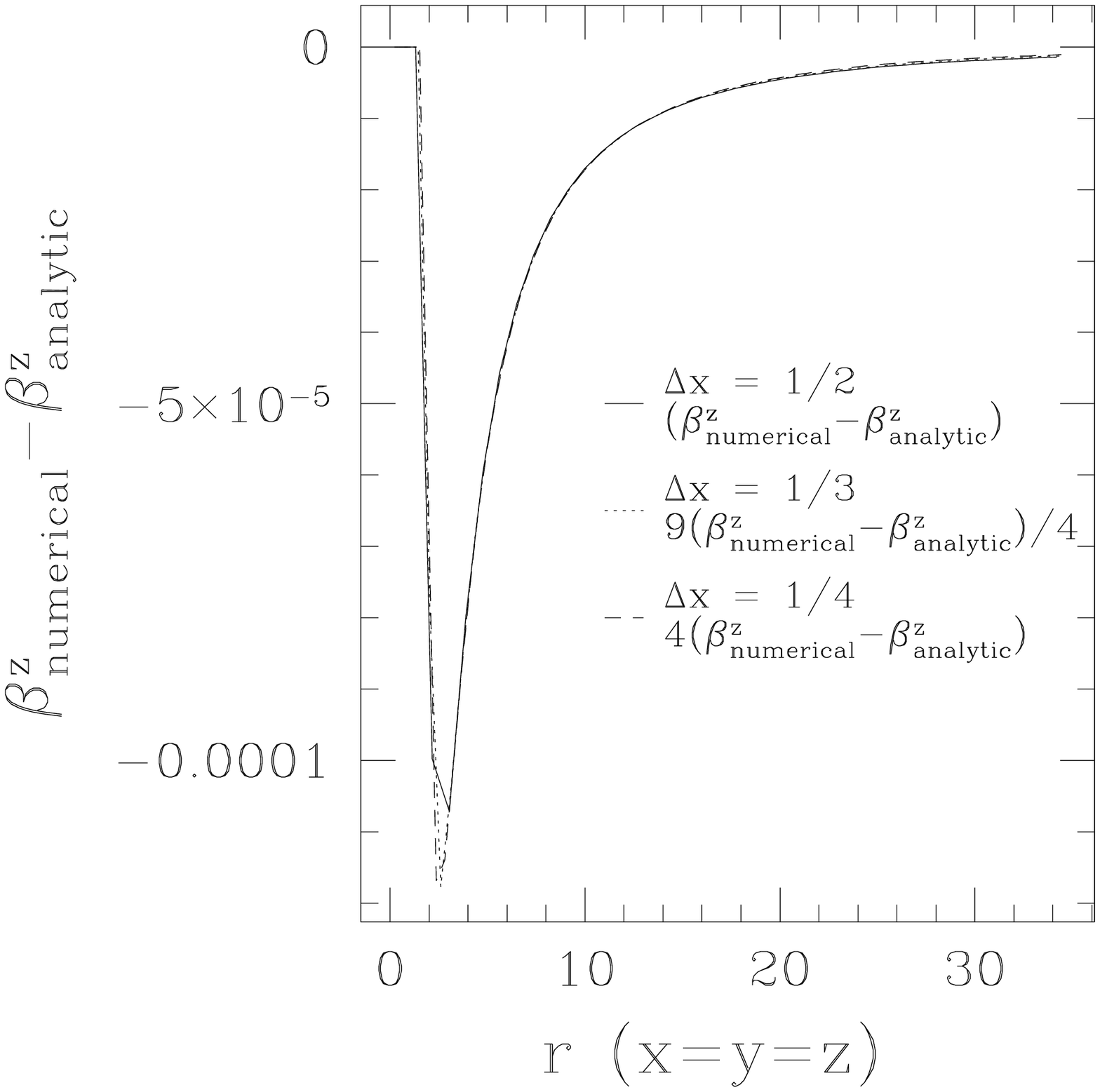}
\end{tabular}
\caption{Numerical errors in the conformal factor $\psi$ and the 
shift $\beta^i$ for three different resolutions $\Delta x$ for a 
single, non-rotating Kerr-Schild black hole.}
\label{singleconv2}
\end{figure*}
  
In this section we present two tests of our code.  The first test
shows second-order convergence to the analytic Kerr-Schild spacetime
for a single black hole.  The second test shows second-order
convergence in a particular (non-equilibrium) binary solution
previously considered by Pfeiffer \cite{pct02}.

\subsubsection{Single Black Hole Tests}

To recover a single non-rotating black hole in Kerr-Schild coordinates
we set $M_B = 0$ and located the black hole $A$ at the origin.  We
then solve equations (\ref{esstart}) -- (\ref{esend})
 imposing Dirichlet inner boundary conditions 
\begin{eqnarray}
\psi & = & 1,  \nonumber \\
\beta^i & = & - \frac{2H l^t l^i}{1+2H l^t l^t}, \label{sbhBC} \\
\alpha & = & \frac{1}{\sqrt{1+2H l^t l^t}} \nonumber
\end{eqnarray}
on a sphere of radius $r_{\rm excision} = 1.8 M_0$, where $M_0$ is the
background mass of one of our black holes at infinite separation.  We
run these tests with the IBM pSeries 690 machine in NCSA. A typical
run with $160\times80\times80$ gridpoints takes about 16 cpu hours and
uses about two giga bytes of memory.  In Fig.~\ref{singleconv2} we
show numerical errors for the conformal factor $\psi$ and the shift
$\beta^i$ for several different resolutions.  The errors scale as
expected, establishing second-order convergence of our code.  We chose
to impose the above Dirichlet condition for $\psi$ instead of the
inner boundary condition (\ref{iibc}) because, in our numerical
implementations, Dirichlet conditions are easier to impose at a fixed
physical location, which is mandatory for achieving second-order
convergence.  For a resolution of $\Delta x/M_0 = 1/6$ and with the
outer boundary at $20M_0$, we find $J/M_0^2 < 10^{-8}$, $M_{\rm
ADM}/M_0 = 0.9990$, and $M_{\rm AH} = 1.0013$, which shows that the
error in our solution is of the order of a fraction of a percent.

\begin{figure*}
\begin{tabular}{rl}
\includegraphics[height=7cm,width=7cm]{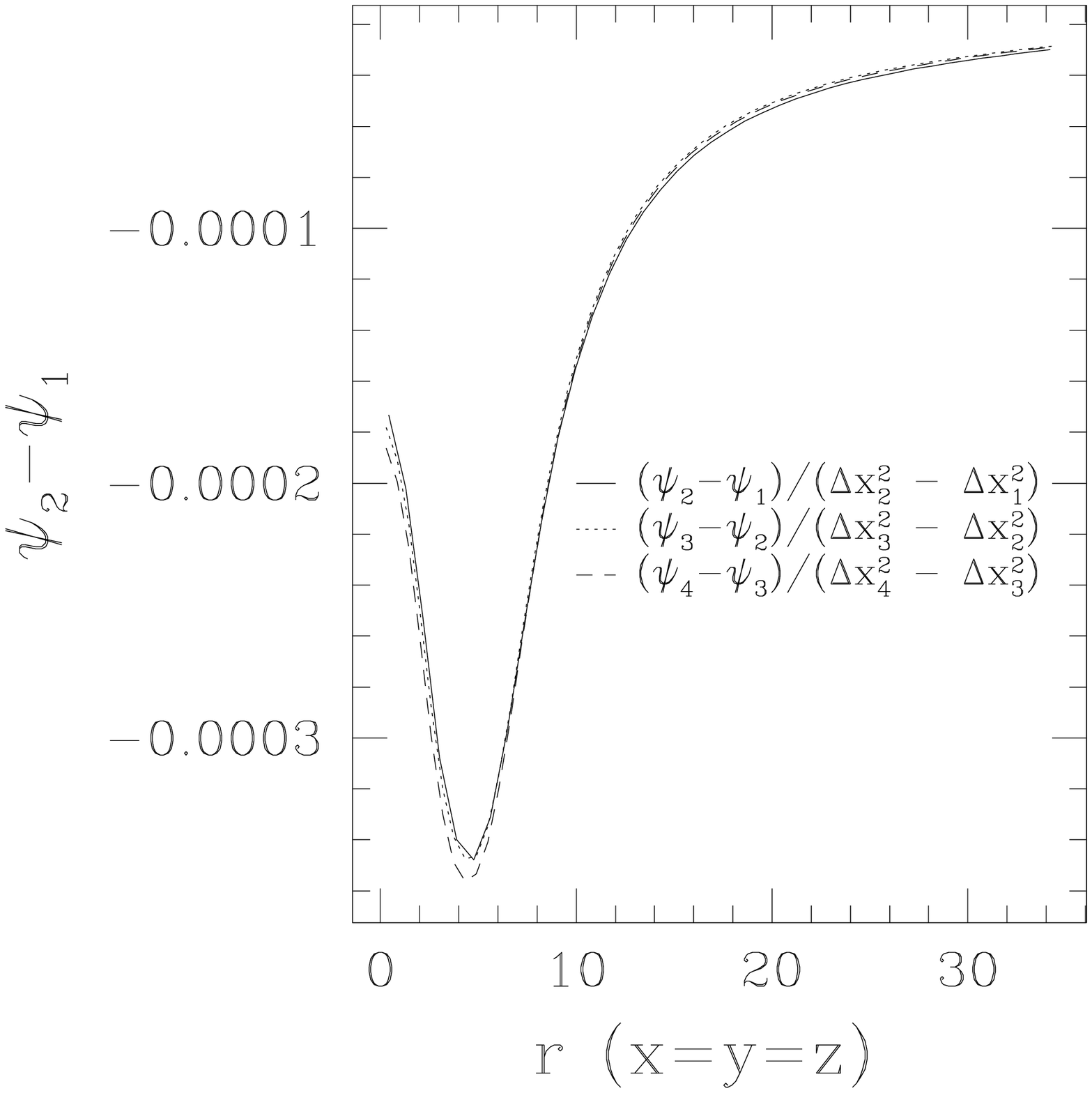}&
\includegraphics[height=7cm,width=7cm]{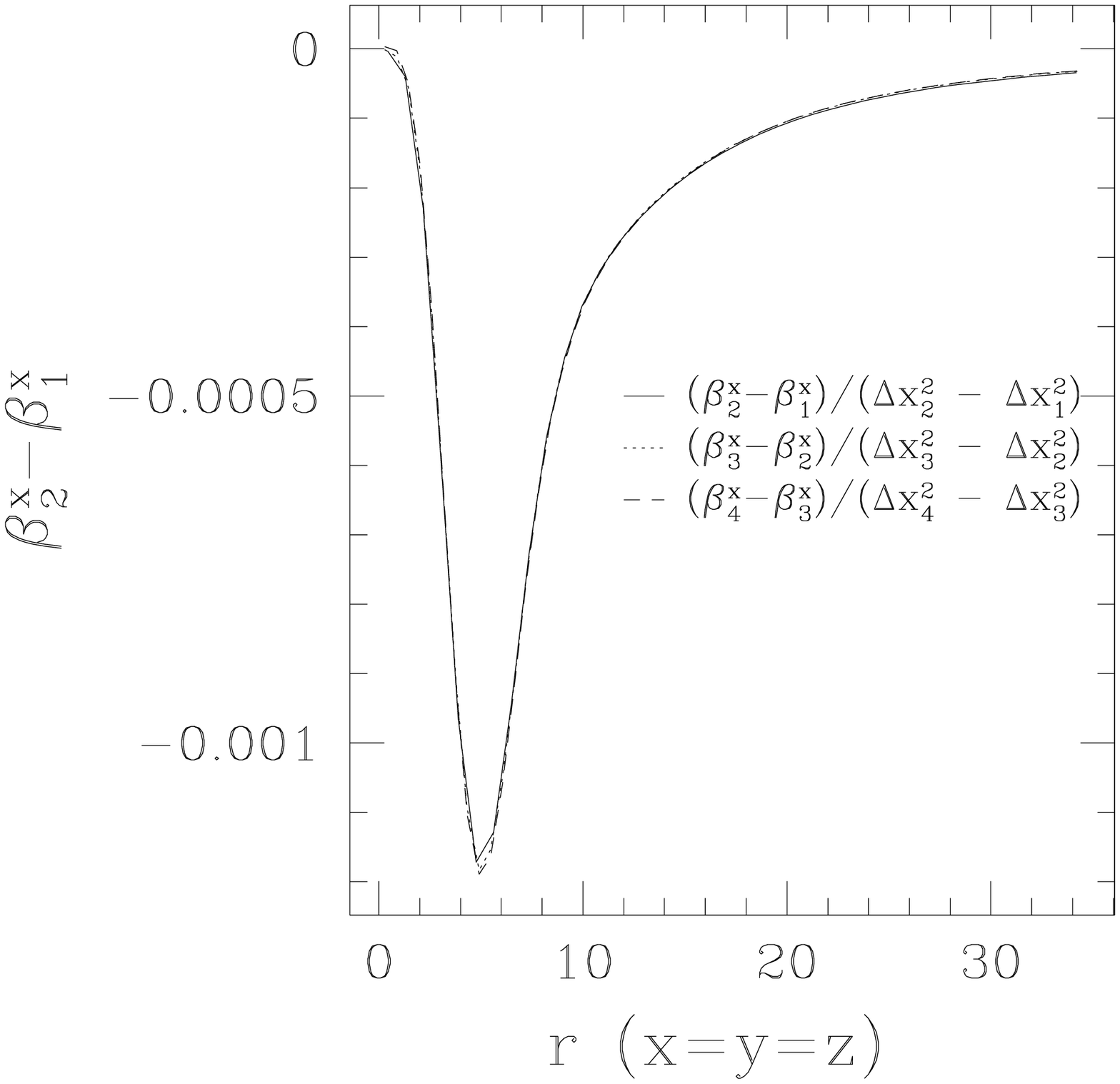} \\
\includegraphics[height=7cm,width=7cm]{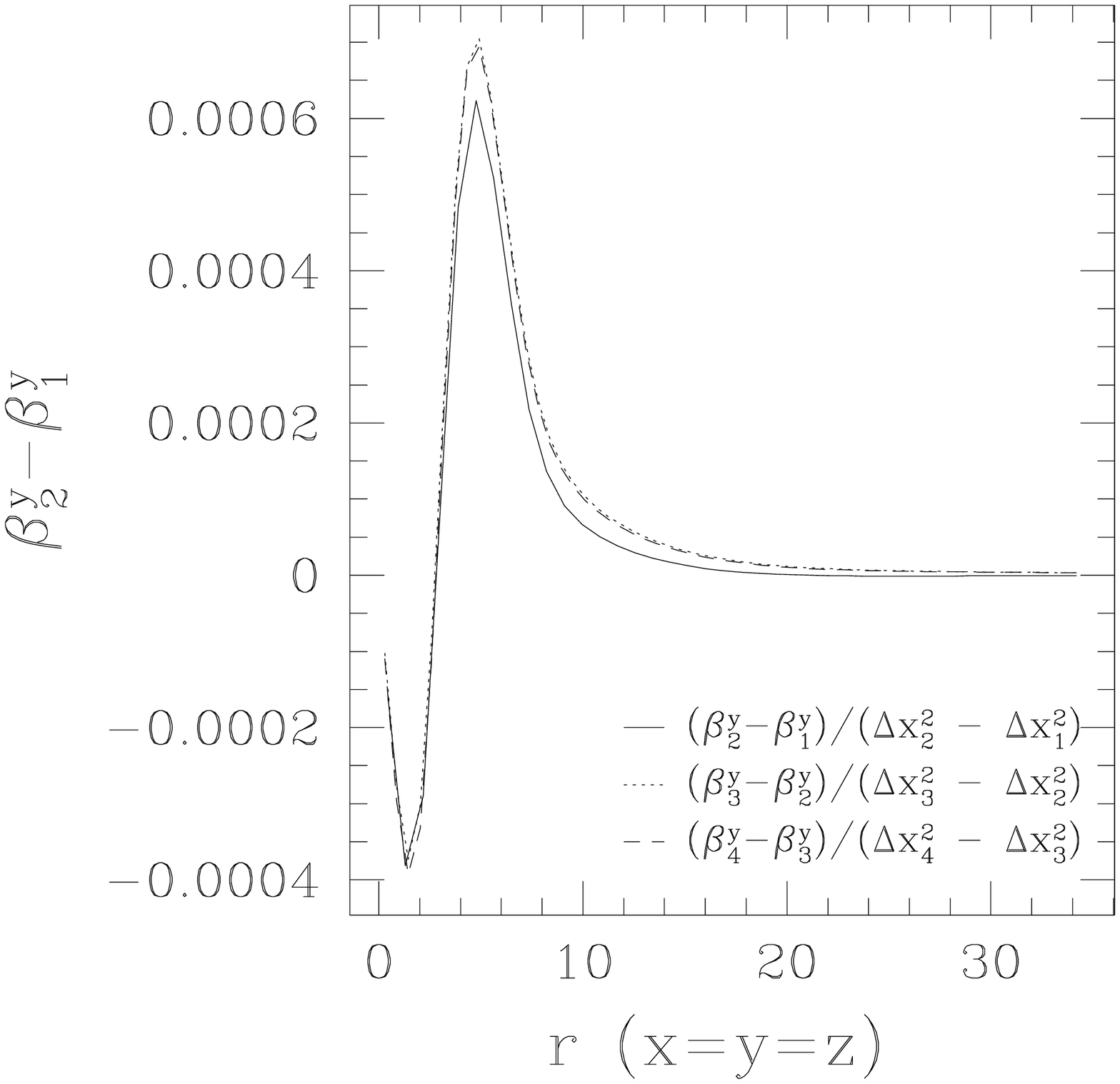}&
\includegraphics[height=7cm,width=7cm]{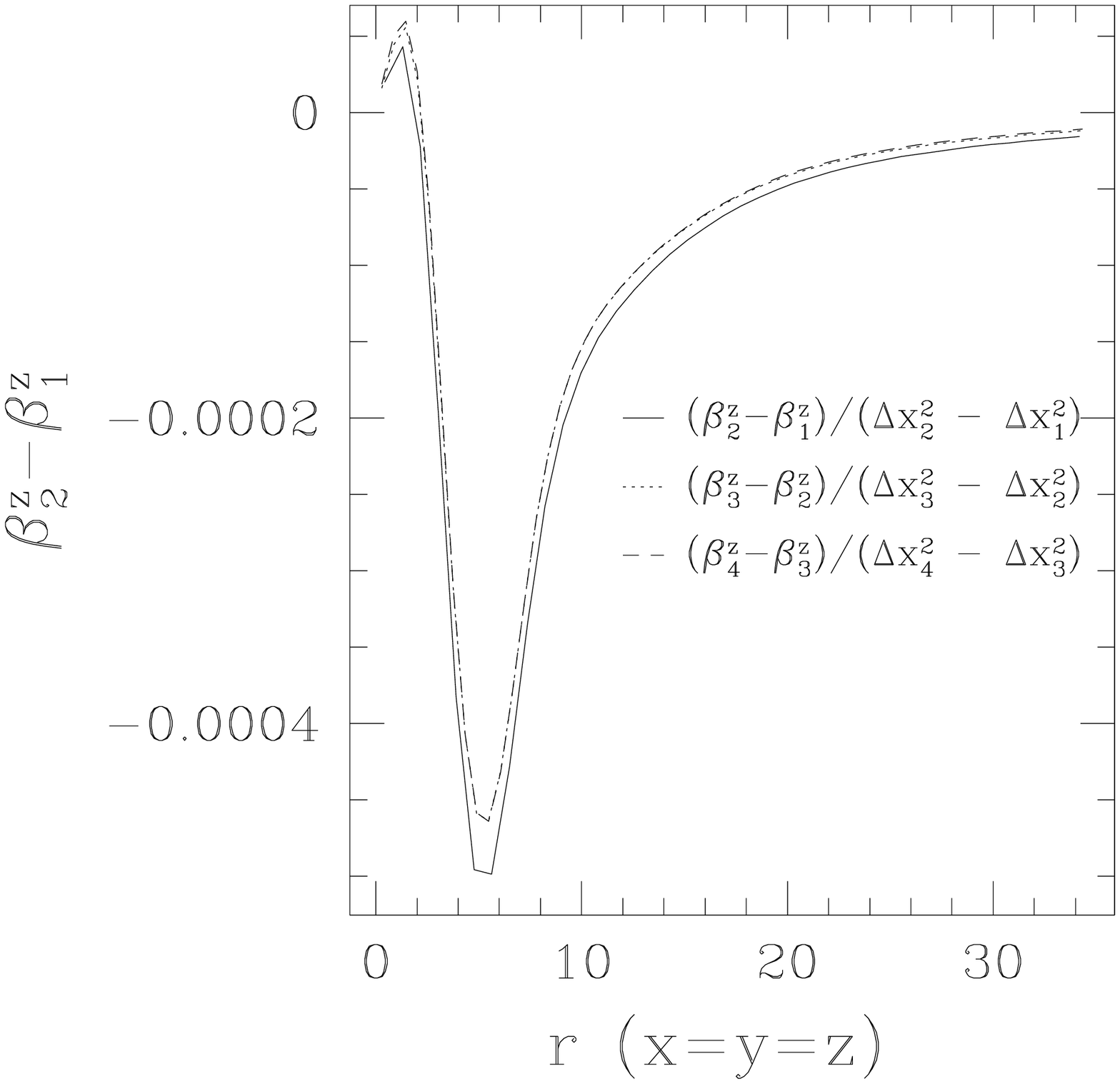}
\end{tabular}
\caption{Numerical convergence of the conformal factor $\psi$ and the
shift $\beta^i$ for the PCT \cite{pct02} binary configuration described
in the text.
We plot the rescaled differences between four different resolutions to
establish second-order convergence (see Appendix \ref{secordconv}).}
\label{binaryconv2}
\end{figure*}

\begin{table}[tb]
\begin{tabular}{|c|c|c|c|c|c|}\hline\hline
$\Delta x/M$&Domain
&$M_{\rm irr}/M$&$M_{\rm ADM}/M$
&$J/M^2$&$\ell/M$\\ \hline
\multicolumn{2}{|c|}{CTS}&1.06528&2.08436&3.3790&10.3971\\ \hline
1/2&small&1.089&2.106&3.389&10.121\\
1/3&small&1.081&2.110&3.391&10.123\\
1/4&small&1.079&2.113&3.391&10.125\\
1/6&small&1.074&2.120&3.392&10.126\\
1/2&big&1.088&2.092&3.379&10.122\\ 1/3&big&1.081&2.093&3.381&10.124\\
1/4&big&1.079&2.094&3.382&10.126\\ 1/6&big&1.074&2.097&3.382&10.127\\
\hline \hline
\end{tabular}
\caption{Comparison of our results with those of Pfeiffer, Cook and
Teukolsky \cite{pct02} for a binary with $M=M_0$, $d=11.75M_0$
and $\Omega=0.0421/M_0$.  Here $M$ is the mass parameter of
both black holes, $M_{\rm irr}$ is the irreducible mass, $M_{\rm ADM}$
is the ADM mass, $J$ is the angular momentum, and $\ell$ is the proper
distance between black hole horizons.  Here a small boundary box
refers to $-20M \leq x \leq 20M$, $0 \leq y \leq 20M$ and $0 \leq z
\leq 20 M$; a large boundary box refers to $-30M \leq x \leq 30M$, $0
\leq y \leq 30M$ and $0 \leq z \leq 30 M$.}
\label{pctcom}
\end{table}

\subsubsection{Comparison with a Previous Binary Black Hole Calculation}
\label{test_binary}

In this Section we compare with numerical results by Pfeiffer, Cook
and Teukolsky \cite{pct02} (hereafter PCT), who used spectral methods
to construct black hole binaries.  PCT solve equations (\ref{esstart})
and (\ref{esend}), but instead of solving (\ref{esmiddle}) for the
lapse they experiment with two different choices for an analytic
densitized lapse function.  Here we compare with their results for
\begin{equation}
\alpha\psi^{-6} = \alpha_A \alpha_B.
\end{equation}
For this test we also adopt their boundary conditions
(\ref{eq:BC-sandwich}).  The implementation of PCT allows for imposing
the outer boundary conditions (\ref{bcstart}) at a much larger
separation than we can afford.  Following PCT we construct a 
nonspinning black hole binary
with background mass $M=M_0$, centers of the excised spheres at
a coordinate separation $d = 11.75M_0$, and angular velocity $\Omega
=0.0421/M_0$. (According to the effective potential method in
\cite{cook1994} this choice of $\Omega$ corresponds to a circular orbit.)

In Fig.~\ref{binaryconv2} we compare our results for the conformal
factor $\psi$ and the shift $\beta^i$ for different resolutions
(all with $r_{\rm excision}=2.0M_0$), and
again establish second-order convergence or our numerical code.  In
Table \ref{pctcom} we tabulate both PCT's and our results for the
binaries irreducible mass, the ADM mass, the angular momentum and the
proper separation between the horizons.  To compute the ADM mass and
the angular momentum we extrapolate our numerical results to a grid
with outer boundaries at $150M_0$ as explained in Appendices
\ref{ADMmass} and \ref{angmom}.  We find values that are very similar
to those found by PCT, but ours do not converge to theirs for a fixed
location of the outer boundary.  We believe that this is caused by the
proximity of our outer boundary, and expect, as the results in Table
\ref{pctcom} suggest, that the agreement would improve by increasing
both the resolution and the distance to the outer boundaries.  From
Table \ref{pctcom} we estimate that the numerical errors in our
results, given the grid resolution and location of the outer
boundaries adopted in this paper, are of the order of about a percent.

\subsection{An approximate inspiral sequence}

\begin{table}
\begin{center}
\begin{tabular}{llll}
\hline
\hline
Reference   & $E_b/\mu$ &$J/(2\mu M_{\rm irr})$&$2\Omega M_{\rm irr}$ \\
\hline
Schwarzschild           & -0.0572       & 3.464         & 0.068 \\
\cite{cook1994}         & -0.09030      & 2.976         & 0.172 \\
\cite{btw00}        	& -0.092        & 2.95          & 0.18  \\
\cite{ggb2} 		& -0.068        & 3.36          & 0.103 \\
\cite{cook03}           & -0.058        & 3.45          & 0.085 \\
This work               & -0.06       	& 3          	& 0.08 \\
\cite{djs2000}  	& -0.0668       & 3.27          & 0.0883\\
\hline
\hline
\end{tabular}
\end{center}
\caption{Values for the binding energy $E_b/\mu$, the angular velocity
$2\Omega M_{\rm irr}$ and the angular momentum $J/(2\mu M_{\rm irr})$
at the ISCO as obtained in different approaches.  The Schwarzschild
results refer to the innermost stable circular orbit of a test
particle in a Schwarzschild spacetime.  Our results for this work are 
determined from the turning point of the binding energy curve in 
Fig.~\ref{Sequence};
but, as discussed in the text, they are prone to large errors.}
\label{ISCO_comp}
\end{table}

\begin{figure*}[tbhp]
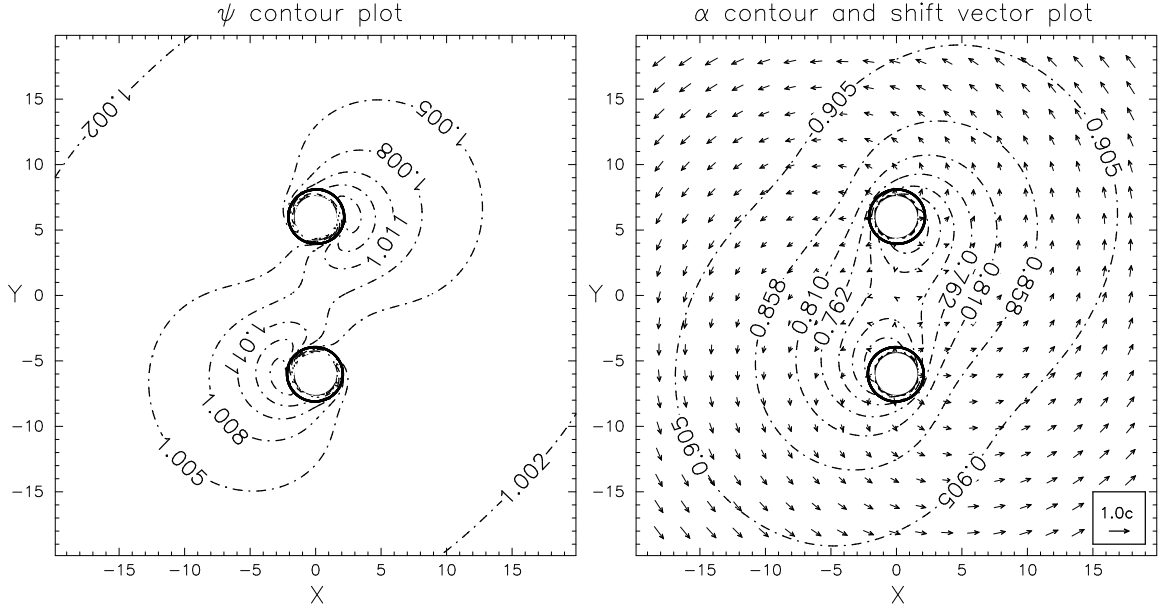

\begin{tabular}{rl}
\includegraphics[width=8cm,angle=270]{fig4a.eps}&
\includegraphics[width=8cm,angle=270]{fig4b.eps}
\end{tabular}
\caption{The contour plots of the binary initial data with coordinate
separation $12M_0$. The corresponding orbital velocity $2\Omega M_{\rm
irr} = 0.0788$.  The left panel shows the contour of the
conformal factor $\psi$; the right panel shows the contour of
lapse $\alpha$ and the shift vector in the equatorial plane. The
excision radius is $1.6 M_0$.  The two thick circles in the plots are
the apparent horizons.  The numerical value of the lapse $\alpha$ on
the apparent horizons exceeds $0.5$ and is therefore positive, as
required for horizon-penetrating coordinates.  It is evident from the
plots that the solutions do not satisfy octant symmetry (see
Appendix \ref{symmetries}).}
\label{Contours}
\end{figure*}

We now proceed to construct an approximate inspiral sequence
for a nonspinning black hole binary system, adopting
the inner boundary conditions described in Section \ref{InnerBC}.
Contours of the conformal factor $\psi$, the lapse $\alpha$ and the
shift $\beta^i$ for one particular binary separation are shown in
Fig.~\ref{Contours}.

\begin{figure*}
\begin{tabular}{rl}
\includegraphics[height=7cm,width=8cm]{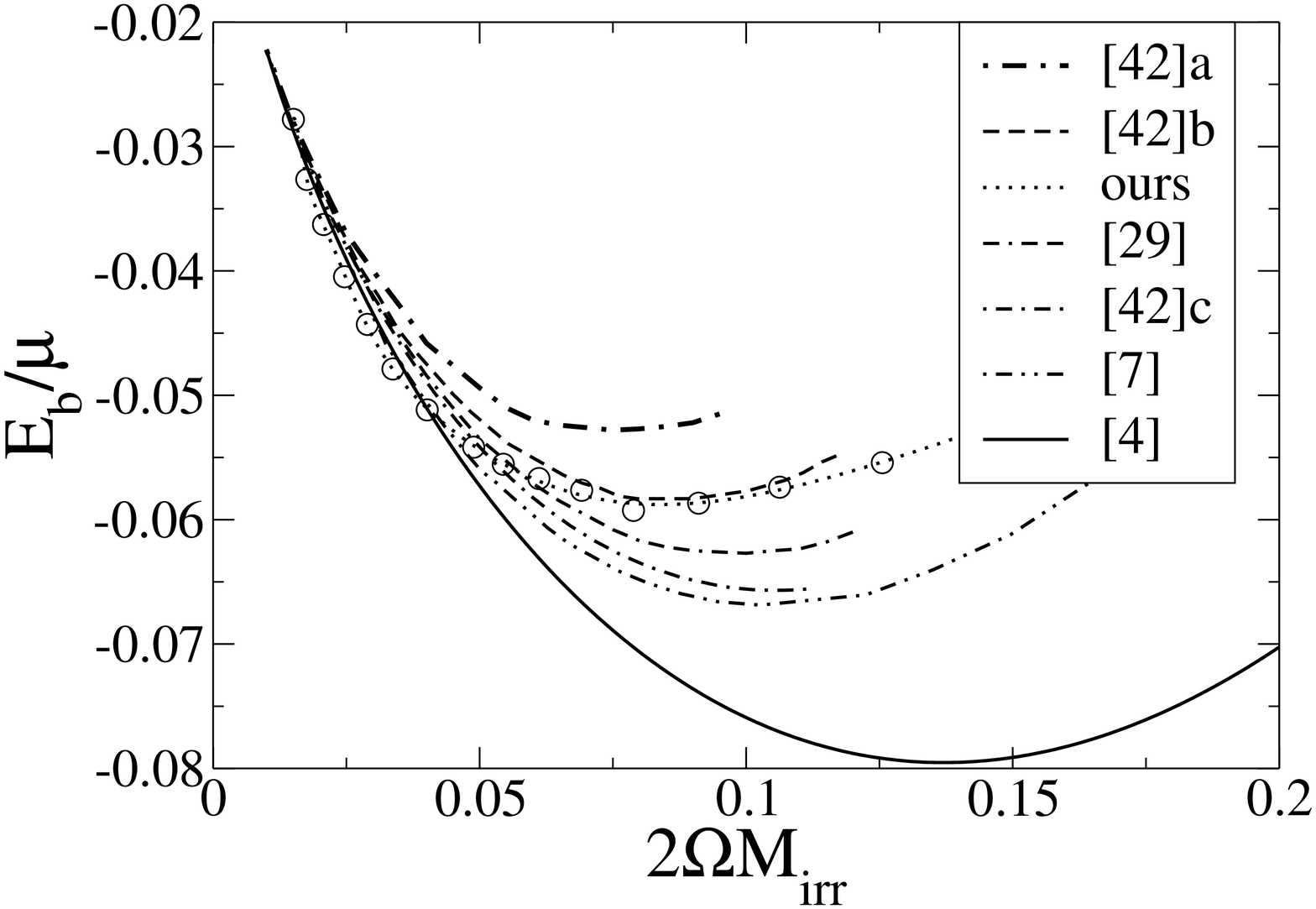}&
\includegraphics[height=7cm,width=8cm]{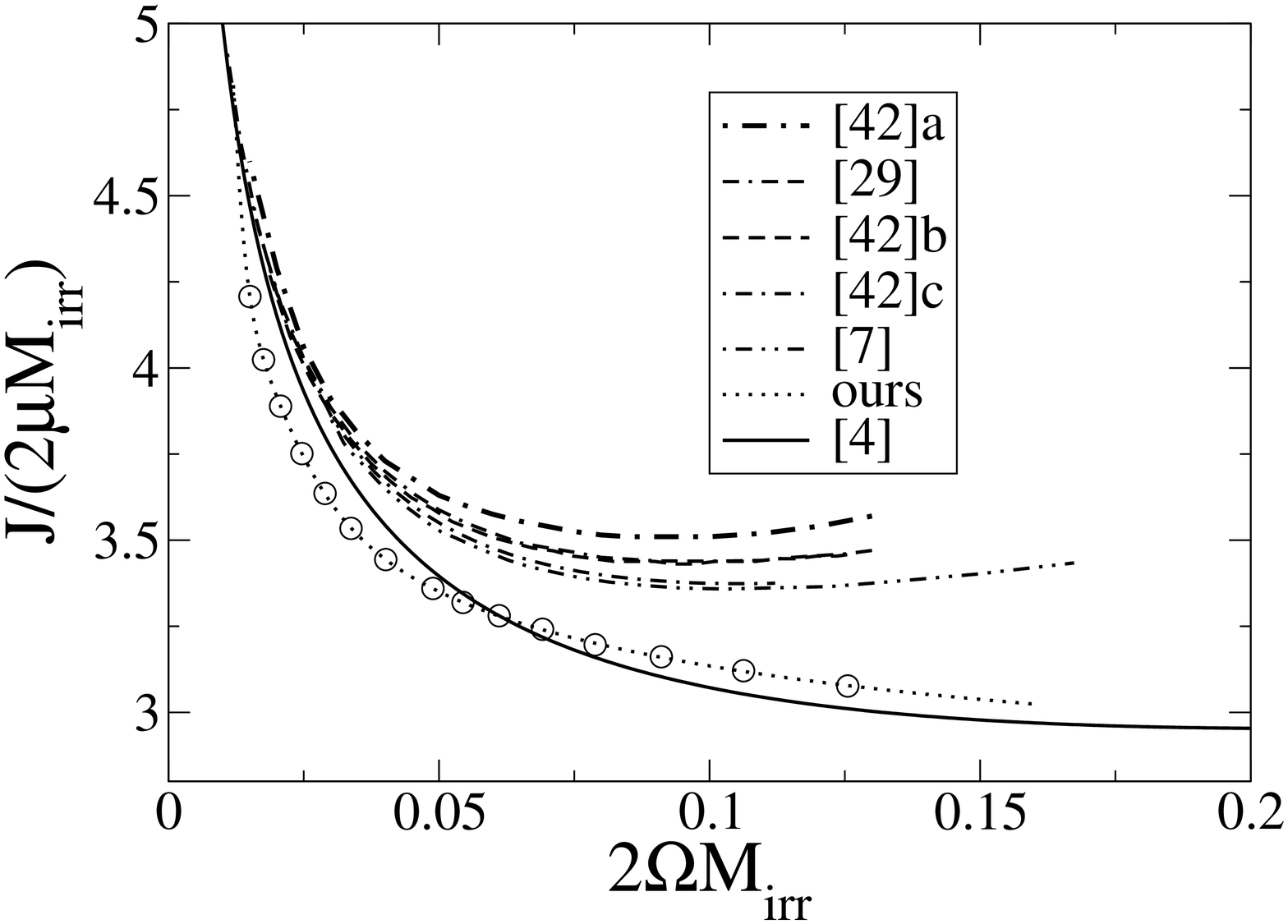}
\end{tabular}
\caption{The binding energy $E_b$ and angular momentum $J$ as a function of
orbital angular velocity.  The dotted lines are least square fits to
our numerical results, marked by open circles.  We compare with thin-sandwich
results in \cite{cook03} for 
an Eddington-Finkelstein slicing and $d\alpha\psi/dr=0$ inner boundary condition
(\cite{cook03}a),
an Eddington-Finkelstein slicing and $\alpha\psi=1/2$ inner boundary condition
(\cite{cook03}b), 
, and a maximal slicing and $d\alpha\psi/dr=\alpha\psi/2r$ inner boundary condition
(\cite{cook03}c),
with the binary initial data in \cite{ggb2},
with the second-order, post-Newtonian sequence in \cite{cook1994},
and with the third-order, post-Newtonian sequence in \cite{dgg02}.
Here $\mu$ is the reduced mass ($M_{\rm irr}/2$)
and $M_{\rm irr}$ is the irreducible mass for one black hole.}
\label{Sequence}
\end{figure*}

The binding energy of an equal mass binary can be defined as
\cite{cook1994}
\begin{equation} \label{E_b}
E_{b} = M_{\rm ADM} - 2 M_{\rm irr}.
\end{equation}
A simultaneous turning point in the binding energy and the angular
momentum locates the ISCO.  In Fig.~\ref{Sequence} we show both the
binding energy and the angular momentum.  We show numerical results
obtained with a resolution of $\Delta x = M_0/4$ and with the outer
boundary at $30 M_0$, which corresponds to the highest resolution and
largest grid run in our comparison in Section \ref{test_binary}.  The
code is run with the IBM pSeries 690 machine in NCSA. Each run with
$240\times120\times120$ gridpoints takes about 1600 cpu hours and uses
about 8 giga bytes of memory.  We set $r_{\rm excision} =1.6M_0$ for
these models. We also compare our results with the numerical results of:
Cook \cite{cook03}, for an Eddington-Finkelstein slicing and $d\alpha\psi/dr=0$
inner boundary condition (\cite{cook03}a),
an Eddington-Finkelstein slicing and $\alpha\psi=1/2$
inner boundary condition (\cite{cook03}b),
and a maximal slicing and $d\alpha\psi/dr=\alpha\psi/2r$
inner boundary condition (\cite{cook03}c);
with the binary initial data in \cite{ggb2};
with the second-order, post-Newtonian sequence in \cite{cook1994};
and with the third-order, post-Newtonian sequence in \cite{dgg02}.

Our results for the binding energy agree fairly well with
those of \cite{cook03}b, while our results for the angular momentum do not.
Note that we find a
turning point in the binding energy, but not in the angular momentum.
There are several possible reasons for these findings.  Solving the
constraints in the thin-sandwich decomposition leads to configurations
that are in approximate equilibrium, but lacking dynamical evolutions,
it is difficult to determine just how good this approximation is 
for this scenario \cite{fn1}.
Another potential reason for our findings are the inner boundary conditions
(\ref{iibc}), which may lead to undesirable deviations from
quasi-equilibrium.  Probably more important, however, is the limited
numerical accuracy of our finite-difference, Cartesian code.
From Table \ref{pctcom} we estimate that
the accuracy of our values for the masses and angular momenta is of
the order of a percent or so.  From equation (\ref{E_b}) the binding
energy is computed as the difference between two masses, and is of the
order of about 10 \% of each of those masses (see Fig.~\ref{Sequence}).
The relative
error in this smaller difference is therefore significantly larger than
the error in each of the terms separately,
and may be as large as 10 \% or more.  Such an
error is large enough to spoil the accuracy of an
inspiral sequence.  However, if taken at face value, the orbital
parameters at the turning point of the binding energy agree fairly
well with recent results for the binary black hole ISCO (see Table
\ref{ISCO_comp}).

While our results are not accurate enough to reliably locate the ISCO,
we do believe that they are suitable for adoption as initial data in
current dynamical evolution calculations in finite difference
implementations. For these purposes, the accuracy of the initial
data only needs to be as small as that of the dynamical evolution
itself.  The individual metric quantities that must be specified as 
initial data are not small differences of large numbers and are determined 
to about a percent.

We also note that solving the constraints in the Bowen-York formalism
leads to higher accuracy solutions even in finite difference
implementations (compare \cite{btw00}).  There, the momentum
constraint can be solved analytically, leaving only the Hamiltonian
constraint to be solved numerically. Moreover, the angular momentum can be
determined analytically in terms of the background quantities.  The
Bowen-York formalism also adopts maximal slicing, $K = 0$, so that
octant symmetry can be adopted and a higher grid resolution can be chosen
(compare Appendix \ref{symmetries}).  In our approach, five coupled
equations are solved simultaneously, and all orbital parameters are
computed numerically from the solutions, which will clearly lead to a
larger numerical error.

\section{Discussion}

We present a method for constructing solutions to the constraint
equations of general relativity, describing quasi-equilibrium 
binary black holes in
nearly circular orbit.  We expect that these solutions are suitable
initial data for dynamical evolution with current finite difference
evolution codes.

We solve the constraint equations in a conformal thin-sandwich
decomposition (e.g.~\cite{york-1999}), and impose quasi-circular orbits
by imposing that the ADM mass of the binary be equal to its Komar mass
(compare \cite{ggb2}).  We adopt a superposition of two Schwarzschild
black holes in Kerr-Schild coordinates as the conformal background
solution.  This background choice leads to horizon-penetrating coordinates,
which are needed for dynamical evolutions, and is likely to produce less
spurious gravitational radiation, at least for rotating black holes.
We present a new set of simple inner boundary conditions, to be imposed
on the excision surface inside the black hole, which we hope leads
to reasonable approximation to equilibrium (compare \cite{cook02}).

We present two numerical tests -- one for the limiting case of an
isolated black hole, and the other for a binary configuration
considered in \cite{pct02}.  We also construct an approximate inspiral
sequence. Our numerical accuracy may not be sufficient to 
track the binding energy to high precision, since it is computed as the small
difference between two significantly larger numbers. However, we do expect that
our solutions provide adequate initial data for current finite-difference
evolution codes.
We also expect that when our formalism is implemented with higher resolution
and/or more accurate numerical schemes, the inspiral sequence may provide 
a more reliable estimate of the ISCO parameters.

\acknowledgments

It is a pleasure to thank P. Marronetti and W. Tichy for useful discussions.
This paper was supported in part by NSF Grants PHY-0090310 and
PHY-0345151 and NASA Grant NNG04GK54G at the University of Illinois
at Urbana-Champaign (UIUC) and by NSF Grant PHY-0139907 at Bowdoin
College.  Numerical calculations were carried out at the National
Center for Supercomputing Applications at UIUC.

\appendix

\section{The ADM Mass Integral}
\label{ADMmass}

%========================================
% Figure
%========================================
\begin{figure}
\includegraphics[height=3.3in]{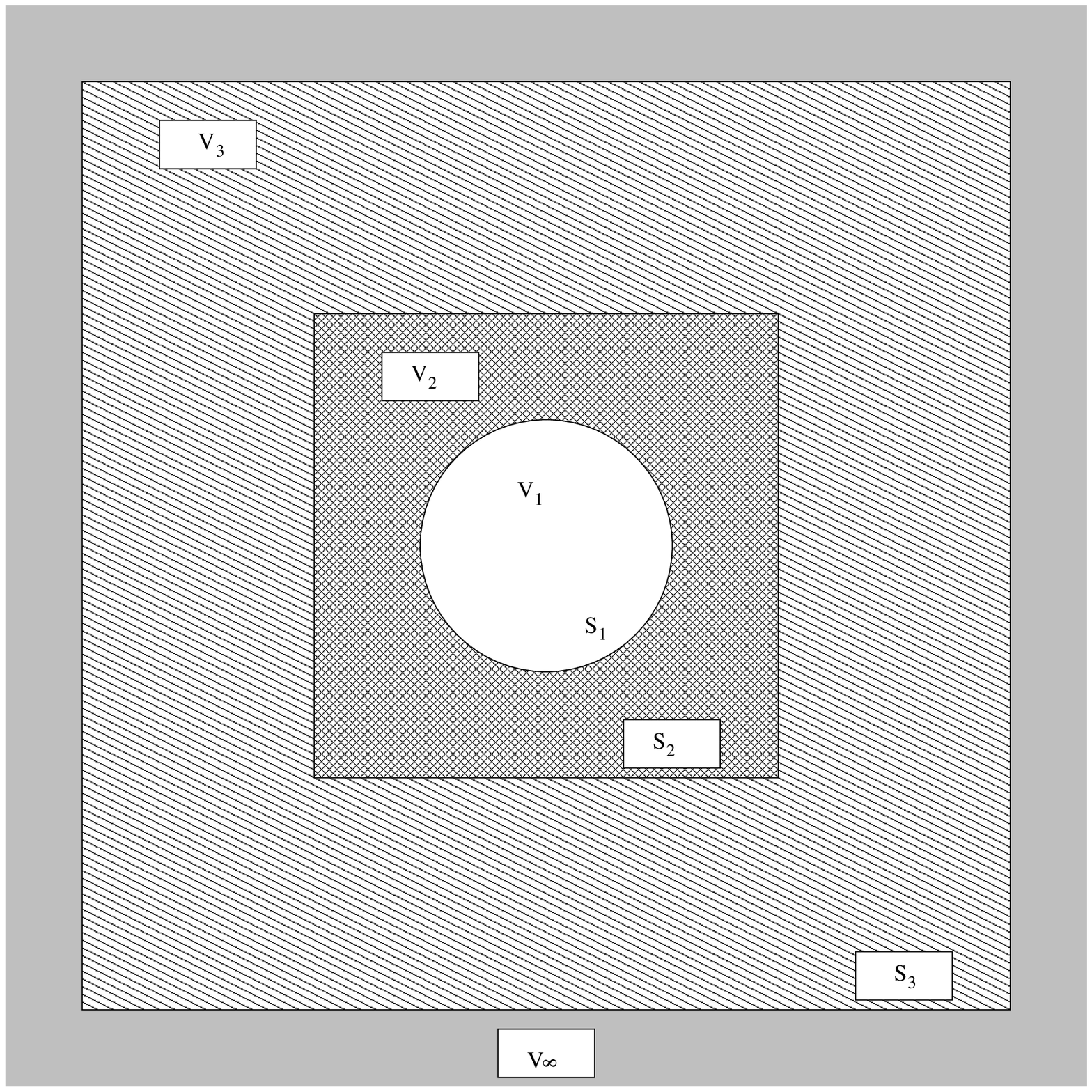}
\caption{
%[HJ: could we include ${\cal S}_\infty$ here?]
The diagram
illustrates the relation between the volumes
$V_1$, $V_2$, and $V_3$ and the surfaces 
${\cal S}_1$, ${\cal S}_2$, and ${\cal S}_3$.  ${\cal S}_1$ is a
boundary outside both black holes, but well inside the computational
domain, ${\cal S}_2$ is a boundary near the outer edge of the
computational domain, and ${\cal S}_3$ is a boundary well outside the
computational domain (at very large, finite radius).}
\label{fig2}
\end{figure}
%========================================

Having conformally decomposed the spatial metric $\gamma_{ij}$, 
the ADM mass integral (\ref{eq:ADM-mass-gen-def}) can be written as
\begin{equation}
 M_{\rm ADM}
   =\frac{1}{16\pi}
    \oint_{{\cal S}_\infty}(\tilde{\Gamma}^i -\tilde{\Gamma}^{ji}{}_j
     - 8\tilde{\nabla}^i\psi)d \tilde{S}_i
\end{equation}
where $\tilde \Gamma^i \equiv \tilde \gamma^{lm} \tilde \Gamma^i_{lm}$
(compare \cite{ybs02}) and where ${\cal S}_\infty$ is a closed surface at
spatial infinity.  For the background data described in Section
\ref{KSbackground} the first two terms can be evaluated analytically
and yield the sum of the two background masses $M_A + M_B$, or, for
equal mass binaries, $2M_0$.  The ADM mass therefore reduces to
\begin{equation}
M_{\rm ADM} = 2 M_0 - \frac{1}{2\pi}
    \oint_{{\cal S}_\infty}\tilde{\nabla}^i\psi d \tilde{S}_i.
\end{equation}
We now convert this surface integral into a volume integral using
Gauss' law; volume integrals are typically more accurate numerically than 
surface integrals. However, since a volume, say $V_1$, containing a black hole
singularity is excised, a surface integral over the surface of that volume,
say ${\cal S}_1$, remains
\begin{eqnarray}
M_{\rm ADM} &=& 2M_0 -\frac{1}{2\pi}\oint_{{\cal S}_\infty}
                \tilde{\nabla}^i\psi d\tilde{S}_i, \nonumber \\
            &=& 2M_0 - \frac{1}{2\pi}\oint_{{\cal S}_1}\tilde{\nabla}^i\psi d\tilde{S}_i
                \nonumber\\
             &&   - \frac{1}{2\pi}\int_{V_2+V_3+V_\infty}
                \sqrt{\tilde{\gamma}}\tilde{\nabla}^2\psi d^3x, \nonumber \\
 &=& 2M_0 - \frac{1}{2\pi}\oint_{{\cal S}_1}\tilde{\nabla}^i\psi d\tilde{S}_i \nonumber\\
 &&         + \frac{1}{16\pi}\int_{V_2+V_3+V_\infty}\sqrt{\tilde{\gamma}}
                [-\psi\tilde{R}\nonumber\\
&& + \psi^{-7}\tilde{A}_{ij}\tilde{A}^{ij} - \frac{2}{3}\psi^5 K^2] d^3x, 
\label{ADM}
\end{eqnarray}
Here we have used the Hamiltonian constraint in the third equality,
and have denoted the volume outside $V_1$ as $V_2+V_3+V_\infty$ as
illustrated in Fig.~\ref{fig2}.  Volume $V_2$ denotes the space
covered by our computational grid.  Given our constraints on numerical
grid resources, this volume extends only to a separation of typically
$30M_0$ from the black holes.  Restricting the ADM
integral (\ref{ADM}) to this volume would introduce a fairly large
error.  We therefore extend the integration to a larger volume $V_3$,
in which the integrand is estimated by extrapolating $\beta^i$,
$\alpha$, and $\psi$ from their values and fall-off conditions on the
outer boundary of the computational grid ${\cal S}_2$: 
\begin{eqnarray}
\psi         &\approx & 1+ \frac{a_1}{r}, \nonumber \\
\alpha       &\approx & 1+ \frac{a_2}{r}, \nonumber \\
\beta^x      &\approx & \bar{\beta}^x + \frac{a_3y}{r^3},\label{falloff} \\
\beta^y      &\approx & \bar{\beta}^y + \frac{a_4x}{r^3}, \nonumber \\
\beta^z      &\approx & \bar{\beta}^z + \frac{a_5z}{r^4}. \nonumber
\end{eqnarray}
Here the $a_i$ are coefficients that are determined as follows.  For
any point in $V_3$, say $\vec r$, we find the intersection of the
position vector $\vec r$ with ${\cal S}_2$.  The value of the function at
that intersection determines the coefficient $a_i$.  Once the
coefficients $a_i$ have been found, the functions $\psi$, $\alpha$ and
$\beta^i$ and hence the integrand of the ADM mass can be evaluated in
$V_3$ (compare \cite{btw00}).  Typically, the boundary of $V_3$ is at
a separation of $150M_0$ from the black holes, so that this
construction increases the volume of our integration by a factor of
125.

\section{The Komar Mass Integral}
\label{KomarMass}               

%========================================
% Figure
%========================================
%\begin{figure}
%\includegraphics[height=2.5in]{fig7.eps}
%\caption{The diagram illustrates the relation between the volume integral
% on the volume $\Omega$ and the surface integrals on the boundaries
% $\partial\Omega_1$ and $\partial\Omega_2$.}
%\label{fig1}
%\end{figure}
%========================================

The Komar mass can be defined for stationary, asymptotically flat spacetimes.
Stationarity implies the existence of a Killing vector $\xi^{\nu}$, which
can be written as 
\begin{equation} \label{xi}
\xi^{\nu} = \alpha n^{\nu} + \beta^{\nu},
\end{equation}
where $n_{\mu} \equiv -\alpha t_{,\mu}$ is the time-like unit normal
on the spatial hypersurfaces $\Sigma_t$.
The Komar mass is defined as
\begin{equation} \label{komar}
M_{\rm K} = -\frac{1}{8\pi} \oint_{{\cal S}_\infty}
\xi^{[\mu;\nu]}dS_{\mu\nu}
=-\frac{1}{4\pi} \oint_{{\cal S}_\infty}\xi^{\mu;\nu}n_{\mu}dS_{\nu}, 
\end{equation}
where ${\cal S}_\infty$ is a closed hypersurface of $\Sigma_t$,
diffeomorphic to a 2-sphere, at spatial infinity, and where we have
used Killing's equation $\xi^{[\mu;\nu]}=\xi^{\mu;\nu}$.  The bi-vector
$dS_{\mu\nu}=2n_{[\mu}dS_{\nu]}$, where $dS_{\nu}$ is a spatial
oriented surface area element, is normal on both ${\cal S}_{\infty}$
and $\Sigma_t$.  From (\ref{xi}) we find
\begin{equation}
\xi^{\mu;\nu}n_{\mu}=-\alpha^{;\nu} +\beta^{\mu;\nu}n_\mu
=-\alpha^{;\nu} -\beta^\mu n_\mu{}^{;\nu}.
\end{equation}
Using the identity $n_\mu{}^{;\nu}=-K_\mu{}^\nu-a_\mu n^\nu$, where 
$a_\mu\equiv n^\nu n_{\mu;\nu}$ is the 4-acceleration of normal observers,
we obtain
\begin{equation}
\xi^{\mu;\nu}n_{\mu}dS_\nu = -(\alpha^{;\mu}-\beta^\nu K_\nu{}^\mu)dS_\mu.
\end{equation}
Inserting this into (\ref{komar}) yields
\begin{equation}\label{komargen}
M_{\rm K} = \frac{1}{4\pi} \oint_{{\cal S}_\infty}
            (\nabla^i\alpha-\beta^jK_j{}^i)dS_i.
\end{equation}
The term $\beta^jK_j{}^i$ often falls off faster than $1/r^2$ in an
asymptotically flat space, in which case its contribution to the
integral vanishes. Here, however, this term must be retained.

The Komar mass is independent of the surface ${\cal S}$ on which the
integral is evaluated, as long as all matter sources are inside of
${\cal S}$.  To demonstrate this, we convert the surface integral in
(\ref{komargen}) into the volume integral
\begin{equation}\label{komarv1}
M_{\rm K} = \frac{1}{4\pi}\int
(\nabla_i\nabla^i\alpha - \beta^j\nabla^iK_{ij} - K_{ij}\nabla^i\beta^j)
\sqrt{\gamma}d^3x.
\end{equation}
This integral can be rewritten by inserting the trace of the evolution 
equation (\ref{eq:K-evolution})
\begin{equation}\label{dda}
\nabla_i\nabla^i\alpha=\alpha\left(K_{ij}K^{ij}+\frac{1}{2}(\rho+s)\right)
 +\beta^i\nabla_iK -\partial_tK,
\end{equation}
and the momentum constraint (\ref{eq:momentum-const})
\begin{equation}\label{dkij}
\nabla^jK_{ij}=\nabla_iK+s_i,
\end{equation}
where, for completeness, we have included the matter sources $\rho$,
$s_i$ and $s$
\begin{eqnarray}
\rho&=&n_\mu n_\nu T^{\mu\nu},\nonumber\\
s_i &=&-\gamma_{i\mu}n_\nu T^{\mu\nu},\\
s &=& \delta^i{}_\mu\gamma_{i\nu}T^{\mu\nu}.\nonumber
\end{eqnarray}
The volume integral (\ref{komarv1}) then becomes
\begin{eqnarray}\label{komarv2}
M_{\rm K} &=& \frac{1}{4\pi}\int \sqrt{\gamma}d^3x
 \Big\{\alpha\left(K_{ij}K^{ij} + \frac{1}{2}(\rho+s)\right)\nonumber\\
 && -\partial_tK - K_{ij}\nabla^i\beta^j - \beta^is_i\Big\}.
\end{eqnarray}
We now use the evolution equation (\ref{eq:g-evolution}) to rewrite
the term $K_{ij}\nabla^i\beta^j$ as
\begin{equation}
K_{ij}\nabla^i\beta^j = \alpha K_{ij}K^{ij}
  + \frac{1}{2}K^{ij}\partial_t\gamma_{ij}.
\end{equation}
This brings the integral (\ref{komarv2}) into the form
\begin{eqnarray}
M_{\rm K} & = & \frac{1}{4\pi}\int \sqrt{\gamma}d^3x
\Big( \frac{1}{2} \alpha(\rho+s) - \beta^is_i \nonumber \\
	& & -\partial_tK
 - \frac{1}{2}K^{ij}\partial_t\gamma_{ij}\Big).  \label{komarv3}
\end{eqnarray}
As in the calculation of the ADM mass, part of the numerical grid may
have to be excluded from the integration, for example if it contains
a black hole singularity.  The integral over an outer surface ${\cal
S}_\infty$ can then be written as a volume integral $V$ and a surface
integral over an inner surface, e.g.~${\cal S}_1$ as in Fig \ref{fig2}
\begin{eqnarray}
M_{\rm K} &=& \frac{1}{4\pi}\oint_{{\cal S}_\infty}
            (\nabla^i\alpha-\beta^jK_j{}^i)dS_i\nonumber\\
 &=&\frac{1}{4\pi}\int_{V_2 + V_3 + V_{\infty}} \sqrt{\gamma}d^3x
\Big(\frac{1}{2}\alpha(\rho+s) - \beta^is_i \nonumber \\ 
& &\qquad\qquad\qquad -\partial_tK
  - \frac{1}{2}K^{ij}\partial_t\gamma_{ij}\Big)\nonumber\\
&&  + \frac{1}{4\pi}\oint_{{\cal S}_1}
\big(\nabla^i\alpha-\beta^jK_j{}^i\big)dS_i.
\end{eqnarray}
From the above assumption of stationarity, the time derivatives of
$\gamma_{ij}$ and $K$ have to vanish, and as long as there are no
matter sources in $\Omega$, $\rho=s=s_i=0$, the volume integral
vanishes and we have
\begin{equation}\label{invkomar}
 \frac{1}{4\pi}\oint_{{\cal S}_1}(\nabla^i\alpha-\beta^jK_j{}^i)dS_i
 = \frac{1}{4\pi}\oint_{{\cal S}_\infty}(\nabla^i\alpha-\beta^jK_j{}^i)dS_i
\end{equation}
(compare \cite{DetS89}).
%[HJ: are you sure this is the right reference?  
%I checked it and didn't see anything about the Komar mass].
       
\section{The Angular Momentum Integral}
\label{angmom}
                            
In Cartesian coordinates, the angular momentum can be defined as
\begin{equation}
%J_i = \frac{\epsilon_{ijk}}{8 \pi} \oint_{{\cal S}_{\infty}} x^j K^{kl} d^2 S_l
J_i = \frac{1}{8\pi}\epsilon_{ij}{}^k\oint_{{\cal S}_{\infty}} x^j K^\ell{}_k d^2 S_\ell
\end{equation}
(see \cite{murchadha74,by1980}).  In this paper we only consider 
rotations about the $z$-axis, and therefore compute only the $z$-component
of the angular momentum which can be rewritten as
\begin{eqnarray} \label{ang_mom}
J_z &=& \frac{1}{8\pi}\oint_{{\cal S}_\infty}(x\tilde{A}^l_{~y}
        - y\tilde{A}^l_{~x})d\tilde{S}_l, \nonumber \\
    &=& \frac{1}{8\pi}\oint_{{\cal S}_1}(x\tilde{A}^l_{~y}
        - y\tilde{A}^l_{~x})d\tilde{S}_l\nonumber\\
        &&+ \frac{1}{8\pi}\int_{V_2+V_3+V_\infty}\Big(\tilde{A}^x_{~y}
        + \frac{2}{3}\psi^6x\tilde{\nabla}_yK  \nonumber \\
    & & - \frac{1}{2}x\tilde{A}_{ij}\partial_y\tilde{\gamma}^{ij}
        - \tilde{A}^y_{~x} - \frac{2}{3}\psi^6y\tilde{\nabla}_xK\nonumber\\
    &&  + \frac{1}{2}y\tilde{A}_{ij}\partial_x\tilde{\gamma}^{ij}\Big)
          \sqrt{\tilde{\gamma}}d^3x. \label{ang}
\end{eqnarray}
As in the calculation of the ADM mass (Appendix \ref{ADMmass}) we have
converted the surface integral into a volume integral for greater
numerical accuracy.  As before, we evaluate the integral from the
numerical data in volume $V_2$ and from extrapolated values in volume
$V_3$.  We neglect only those contributions to the integral from
volume $V_{\infty}$.

\section{Inertial and Rotating Frames}

Rotating frames are not asymptotically flat, so that the expressions
for the ADM (Appendix \ref{ADMmass}), angular momentum (Appendix
\ref{angmom}) and Komar mass (Appendix \ref{KomarMass}) have to be
re-evaluated.

The barred coordinates $\bar{t}$, $\bar{x}$, $\bar{y}$ and $\bar{z}$
in an inertial frame are related to the unbarred coordinates $t$,
$x$, $y$ and $z$ in a rotating frame by the transformation
\begin{equation}
\begin{array}{rcl}
\bar{t}&=&t,\\
\bar{x}&=&x\cos(\omega t)-y\sin(\omega t),\\
\bar{y}&=&x\sin(\omega t)+y\cos(\omega t),\\
\bar{z}&=&z.
\end{array}
\end{equation}
Here we are assuming a constant angular velocity $\vec \omega =
(0,0,\omega)$ and rotation about the $z$-axis.  At an arbitrary instant
$\bar{t}=t=0$ at which the two frames are aligned the gravitational
field variables are related by \cite{dmsb03}
\begin{eqnarray}
\alpha&=&\bar{\alpha},\nonumber\\
\beta^i&=&\bar{\beta}^i+(\vec{\omega}\times\vec{r})^i,\nonumber\\
\gamma_{ij}&=&\bar{\gamma}_{ij},\\
K_{ij}&=&\bar{K}_{ij},\nonumber
\end{eqnarray}
The only effect of this transformation is therefore the appearance of a new
term in the shift.  Since the shift does not enter the integrals for the 
ADM mass nor the angular momentum, those quantities remain unchanged and
we only have to re-evaluate the Komar mass.

Transforming between the rotating and inertial frame, we find that the 
Komar mass in the rotating frame $M_K$ is related to that in the 
inertial frame $\bar M_K$ by
\begin{eqnarray}
M_K  & = & \frac{1}{4\pi} \oint(\nabla^i\alpha-\beta^jK_j{}^i)dS_i
  \nonumber\\
 & = &\frac{1}{4\pi}\oint (\nabla^i\bar{\alpha}
-\bar{\beta}^j\bar{K}_j{}^i)d\bar{S}_i -
\frac{1}{4\pi}\oint(\vec{\omega}\times\vec{r})^j
\bar{K}_j{}^id\bar{S}_i\nonumber\\
& = & \bar M_{\rm K} - 
	\frac{\omega}{4\pi}\epsilon^j{}_{z\ell}
\oint x^\ell\bar{K}_j{}^id\bar{S}_i\nonumber\\
& = & \bar M_{\rm K} - 2 \omega J\label{m2wj}
\end{eqnarray}
(note that the angular momentum is the same in both frames, so $\bar J
= J$).

\section{Symmetries of the Computational Domain}
\label{symmetries}

Symmetries can be used to reduce the size of the computational grid,
making it desirable to incorporate as many symmetries as possible.  One
might expect that the binary black hole configuration studied in this
paper would allow for octant symmetry.  In this Appendix we show that
this is not the case, due to the presence of a non-zero trace of the
extrinsic curvature (i.e.~non-maximal slicing).

Consider, for example, the momentum constraint
(\ref{eq:momentum-const}) and assume, for simplicity, conformal
flatness.  On the $x=0$ plane, it would be natural to assume that
$\beta^x$ be symmetric, while $\beta^y$ and $\beta^z$ be
anti-symmetric (compare, for example, \cite{twbNS}).  Computing
$\tilde A^{ij}$ from the shift according to (\ref{eq:TS-conf-gdot}),
assuming that all scalars are symmetric on all coordinate planes,
shows that $\tilde A^{xx}$, $\tilde A^{yy}$, $\tilde A^{yz}$ and
$\tilde A^{zz}$ are all anti-symmetric on the $x=0$ plane, while
$\tilde A^{xy}$ and $\tilde A^{xz}$ are symmetric.  The divergence
$\nabla_i \tilde A^{xi}$, for example, is then symmetric on the $x=0$
plane.  The gradient $\nabla^i K$, however, must be anti-symmetric,
meaning that the momentum constraint (\ref{eq:momentum-const})
violates this symmetry assumption.  Similar arguments hold on the
$y=0$ plane.  Under the assumption of maximal slicing $K=0$, octant
symmetry can be adopted, but in this paper we adopt a Kerr-Schild
background with $K \ne 0$.  The above issue does not apply on the
$z=0$ plane, so that equatorial symmetry can be assumed even in the
non-maximal slicing case $K \ne 0$.

It is possible, however, to adopt $\pi$-symmetry, whereby
\begin{equation}
f(-x,-y,z)=\sigma f(x,y,z).
\end{equation}
For scalar functions we have $\sigma = 1$, while for the $x$, $y$ and
$z$-component of the shift we have $\sigma_x=-1$, $\sigma_y=-1$, and
$\sigma_z=1$, respectively.

\section{Second-Order Convergence} 
\label{secordconv}

Second-order convergence is most easily demonstrated by doubling the 
computational grid resolution several times and showing that numerical
errors scale in the expected way.  Given the constraints of computational
resources it is often impossible to double the grid size several times,
so instead we establish second-order convergence by considering three
arbitrary (but different) grid spacings $h_1$, $h_2$, and $h_3$.

Let $Q(h)$ be a quantity obtained from a finite difference scheme
with spacing $h$.  A Taylor expansion around $h = 0$ yields
\begin{eqnarray}
  Q(h) &=& Q(0) + h\left.\frac{\partial Q}{\partial h} \right|_{h=0}
 + \frac{h^2}{2}\left. \frac{\partial^2 Q}{\partial h^2}Q \right|_{h=0}
           + O(h^3) \nonumber \\
  & \equiv & Q_0 + Q_1 h + Q_2 h^2 + O(h^3). \label{expansion}
\end{eqnarray}
Second-order convergence implies that $Q_1 = 0$.  Given two different 
resolutions $h_1$ and $h_2$ we can eliminate $Q_0$ and find
\begin{equation}
Q_2 = \frac{Q(h_2)-Q(h_1)}{h_2^2-h_1^2} + O(h).
\end{equation}
Alternatively, $Q_2$ can be computed from the two grid spacings $h_2$
and $h_3$.  Second-order convergence can then be established by
showing that the differences between different values for $Q_2$
decrease at least as fast as $h$.

%========================================


\begin{thebibliography}{99}
\bibitem{NoScale} In this paper we do not employ scaling to avoid the 
iteration for $M_{\rm irr}$ as for
example in eqns (79)-(82) in \cite{ggb2}.
Unlike \cite{ggb2,cook03} we do not impose the inner
boundary conditions on the apparent horizon, so that scaling would
change the location of the excision radius with respect to the
apparent horizon.  This behavior would complicate testing for
convergence, as well as comparisons of data from different runs, so
that we decided not to use scaling for the calculations presented
in this paper.
\end{thebibliography}
\end{document}